\documentclass[amsmath,amssymb,aps,11pt,showkeys]{revtex4-2}

\usepackage{xcolor}
\usepackage{stix}
\usepackage{amssymb}
\usepackage{graphicx}
\usepackage{dcolumn}
\usepackage{bm}
\usepackage{setspace}
\usepackage{acronym}
\usepackage{babel}
\usepackage{algpseudocode}
\usepackage{comment}
\usepackage{hyperref}

\usepackage{scalerel}
\usepackage{tikz}
\usetikzlibrary{svg.path}

\definecolor{orcidlogocol}{HTML}{A6CE39}
\tikzset{
  orcidlogo/.pic={
    \fill[orcidlogocol] svg{M256,128c0,70.7-57.3,128-128,128C57.3,256,0,198.7,0,128C0,57.3,57.3,0,128,0C198.7,0,256,57.3,256,128z};
    \fill[white] svg{M86.3,186.2H70.9V79.1h15.4v48.4V186.2z}
                 svg{M108.9,79.1h41.6c39.6,0,57,28.3,57,53.6c0,27.5-21.5,53.6-56.8,53.6h-41.8V79.1z M124.3,172.4h24.5c34.9,0,42.9-26.5,42.9-39.7c0-21.5-13.7-39.7-43.7-39.7h-23.7V172.4z}
                 svg{M88.7,56.8c0,5.5-4.5,10.1-10.1,10.1c-5.6,0-10.1-4.6-10.1-10.1c0-5.6,4.5-10.1,10.1-10.1C84.2,46.7,88.7,51.3,88.7,56.8z};
  }
}

\newcommand\orcidicon[1]{\href{https://orcid.org/#1}{\mbox{\scalerel*{
\begin{tikzpicture}[yscale=-1,transform shape]
\pic{orcidlogo};
\end{tikzpicture}
}{|}}}}

\makeatletter
\newcounter{algorithm}
\renewcommand{\thealgorithm}{\arabic{algorithm}}

\newenvironment{algorithm}[1][tbp]
{%
  \refstepcounter{algorithm}%
  \par\medskip
  \noindent\begin{minipage}{\linewidth}
  \hrule\medskip
  \noindent\textbf{Algorithm~\thealgorithm.}%
  \def\@captype{algorithm}%
}
{%
  \medskip\hrule
  \end{minipage}
  \par\medskip
}
\makeatother

\acrodef{SWBLI}{shock wave-boundary layer interaction}
\acrodef{TSM}{Time Spectral Method}
\acrodef{STSM}{Space-Time Spectral Method}
\acrodef{CFD}{Computational Fluid Dynamics}
\acrodef{DNS}{direct numerical simulation}
\acrodef{LES}{large eddy simulation}
\acrodef{PIV}{particle image velocimetry}
\acrodef{APG}{adverse pressure gradient}
\acrodef{ZPG}{zero pressure gradient}
\acrodef{WNL}{weakly non-linear}
\acrodef{N-S}{Navier-Stokes}
\acrodef{T-S}{Tollmien-Schlichting}
\acrodef{HBNS}{Harmonic-balanced Navier-Stokes}
\acrodef{AHBM}{Analytical Harmonic Balance Method}
\acrodef{DFT}{Discrete Fourier Transform}
\acrodef{IDFT}{Inverse Discrete Fourier Transform}
\acrodef{FE MUSCL}{Flux-Extrapolated MUSCL}
\acrodef{AD}{Algorithmic Differentiation}
\acrodef{GMRES}{Generalized Minimal Residual method}
\acrodef{PSE}{parabolized stability equations}
\acrodef{RA}{resolvent analysis}
\acrodef{GSA}{global stability analysis}
\acrodef{LST}{linear stability theory}
\acrodef{Pyst}{Python Spectral Time}
\acrodef{3D}{three-dimensional}
\acrodef{2D}{two-dimensional}

\begin{document}

\preprint{APS/123-QED}

\title{Optimal non-linear mechanisms for laminar--turbulent transition\\of a shock-induced separated shear layer}

\author{Flavio Savarino}
 \email{flavio.savarino17@imperial.ac.uk}
\affiliation{%
 Department of Aeronautics, Imperial College London\\
 Exhibition Rd, London SW7 2AZ, UK}%
  
\author{Denis Sipp}
 \email{denis.sipp@onera.fr}
\affiliation{%
 DAAA, ONERA, Université Paris-Saclay\\
 8 rue des Vertugadins, 92190, Meudon, France}%

\author{Georgios Rigas}
 \email{g.rigas@imperial.ac.uk}
\affiliation{%
 Department of Aeronautics, Imperial College London\\
 Exhibition Rd, London SW7 2AZ, UK
}%

\date{\today}

\pagebreak
\begin{abstract}
Laminar–turbulent transition in shock wave–boundary-layer interactions (SWBLI) remains a critical challenge for hypersonic vehicle design, with strong implications for drag, heat transfer, and structural loads. Linear optimal perturbation analyses can isolate candidate instabilities, but identifying the full route to breakdown in SWBLI requires nonlinear optimisation.
In this study, we characterise the optimal transition pathway in a globally stable yet convectively unstable Mach 2.15 oblique SWBLI using a nonlinear input–output optimisation framework based on the space–time spectral Navier–Stokes formulation (Poulain et al., Comput. Fluids, 2024). The nonlinear frequency-domain approach captures mean-flow distortion, resolves triadic energy transfers, and extracts intrinsic nonlinear stresses that activate additional instability mechanisms and ultimately lead to breakdown.
We identify an efficient four-stage transition pathway: (1) optimal forcing of oblique first Mack mode waves at moderate frequencies; (2) non-linear self-interaction of counter-propagating Mack waves generating streamwise Görtler-like vortices in the reattachment region where streamline curvature peaks; (3) lift-up of streamwise velocity streaks by these vortices; and (4) sub-harmonic sinuous secondary instability leading to streak breakdown. Optimization across forcing amplitudes from infinitesimal to transitional levels yields quasi-invariant optimal forcing structures, demonstrating that exciting the oblique first Mack mode alone suffices to trigger the entire turbulent cascade. Parametric studies spanning frequency-wavenumber space and forcing configurations confirm this preferential pathway. 
By resolving non-linear energy transfers through a finite number of harmonics, this work establishes a computationally tractable framework for transition prediction and control strategy development in high-speed separated flows, bridging the gap between linear stability theory and fully turbulent simulation.

\end{abstract}


\maketitle


\section{\label{sec:introduction}Introduction}

High-speed aerospace systems, ranging from supersonic aircraft and air-breathing engine inlets to re-entry vehicles, face severe aerodynamic challenges arising from the interaction between shock waves and boundary layers \cite{babinsky_harvey_2011}. In \ac{SWBLI}s, an incident shock imposes a sudden \ac{APG} on the near-wall flow, often inducing separation and reattachment and activating multiple instability mechanisms that strongly affect drag, heat transfer, and overall aerothermal performance \cite{smiths_dussauge_2006,clemens2014low,piponniau2009simple,pirozzoli_bernardini_grasso_2010,touber2011low}. These interactions therefore influence aerodynamic efficiency, thermal loads, structural integrity, and the design of thermal protection systems, and they shape the requirements and opportunities for flow-control strategies in high-Mach-number applications \cite{vega_gramola2020}. Improving our ability to understand, characterise, and predict SWBLIs is thus central to the design of next-generation hypersonic vehicles.

 In configurations where the boundary layer upstream of the impinging shock is laminar, the imposed \ac{APG} can then trigger laminar--turbulent transition of the separated shear layer, a process that increases skin friction and heat transfer and alters the distribution of aerodynamic loads \cite{sandham_schülein_wagner_willems_steelant_2014,zuo_memmolo_huang_pirozzoli_2019,cao_hao_klioutchnikov_wen_olivier_heufer_2022}. While fully turbulent \ac{SWBLI}s have been studied extensively, transitional interactions remain comparatively less understood \cite{robinet2007bifurcations,Sansica_Sandham_Hu_2016}. In particular, the route from receptivity and linear instability \cite{bugeat_robinet_chassaing_sagaut_2022} to breakdown in laminar \ac{SWBLI}s---and the associated sequence and coupling of instability mechanisms---still challenge predictive models. Clarifying this transition pathway is the aim of the present study. 


\subsection{Linear instability mechanisms}\label{subsec:linear instability mechanisms SWBLI}
Transition in \ac{SWBLI}s can involve a range of instability mechanisms, including absolute (self-excited) and convective instabilities \cite{piponniau2009simple,clemens2014low,Sansica_Sandham_Hu_2016,Dwivedi2019,fang_zheltovodov_yao_moulinec_emerson_2020,lugrin_beneddine_leclercq_garnier_bur_2021,cao_hao_klioutchnikov_olivier_wen_2021,sawant_theofilis_levin_2022,dwivedi_sidharth_jovanovic_2022,hao_2023,Song_Hao_2025}. Linear stability analyses, such as \ac{GSA} and \ac{RA}, have been widely used to characterise disturbance growth by linearising the compressible \ac{N-S} equations about either a base flow or a time-averaged flow, depending on the state of the boundary layer.

\ac{GSA} has shown that the separation bubble may support an isolated, weakly \ac{3D} eigenmode that becomes temporally unstable when the shock-induced \ac{APG} exceeds a threshold \cite{robinet2007bifurcations,guiho_alizard_robinet_2016,sidharth2018,cao_hao_klioutchnikov_wen_olivier_heufer_2022,hao_2023,song2023}. One interpretation is the shear-layer/separated-zone model of \cite{piponniau2009simple}, in which coherent shear-layer structures are hypothesised to drive a periodic expansion–contraction (``breathing’’) of the bubble. This breathing can couple to the shock/bubble system, producing low-frequency unsteadiness that modulates the effective pressure gradient seen by the boundary layer (sometimes termed buffet-like unsteadiness). Typical frequencies of order $St_{L_{\mathrm{sep}}}\sim10^{-2}$ (based on the separation length $L_{\mathrm{sep}}$) have been reported \cite{gaitonde2015progress,guiho_alizard_robinet_2016,rabey_jammy_bruce_sandham_2019,Threadgill2020,cao_hao_klioutchnikov_olivier_wen_2021,bugeat_robinet_chassaing_sagaut_2022,hao_2023}, i.e. roughly two orders of magnitude lower than the dominant shear-layer instability frequencies. An alternative explanation proposed by \cite{touber2011low} is that the shock foot behaves as a low-pass filter to incoming disturbances, selectively amplifying the low-frequency part of the spectrum. Later, \cite{clemens2014low,Hildebrand2018} suggested that this unsteadiness results from an interplay between the \ac{APG}, flow non-parallelism, and shock corrugation. Despite substantial effort, a general consensus on the dominant mechanism has not been established. Beyond its role in low-frequency shock/bubble unsteadiness and structural fatigue, this mode may also participate in transition. In particular, \cite{song2023} examined the nonlinear evolution of this mode in a globally unstable laminar \ac{SWBLI} using \ac{DNS} and found a route to turbulence that does not require external forcing, but is initiated by nonlinear growth of the primary bubble mode. A complete picture of global-instability-driven transition in \ac{SWBLI}s nevertheless remains unavailable.

 Complementing bubble-related global modes, compressible boundary layers are also susceptible to the first and second Mack modes. The first Mack mode is a shear-driven viscous instability, whereas the second Mack mode is a higher-frequency, predominantly \ac{2D} acoustic mode that is trapped within the boundary layer. These mechanisms have been studied using local \ac{LST} \cite{Sansica_Sandham_Hu_2016,Khotyanovsky2019}, global resolvent approaches \cite{Dwivedi2020PRF,bugeat_robinet_chassaing_sagaut_2022,Song_Hao_2025}, and \ac{PSE} \cite{li_malik_1995,li_choudhari_paredes_2022}, revealing unstable frequency--wavenumber bands that are central to transition in high-speed flows. Both modes have been reported in supersonic and hypersonic \ac{SWBLI}s \cite{Sansica_Sandham_Hu_2016,kuehl_paredes_2016,bugeat_robinet_chassaing_sagaut_2022,Song_Hao_2025}. In particular, the first Mack mode, excited by free-stream oblique wave-like disturbances at frequencies $St_{L_{\mathrm{sep}}}\sim10^{-1}$--$10^{0}$, has been identified as an efficient trigger for transition of the separated shear layer \cite{Sansica_Sandham_Hu_2016,Dwivedi2020,lugrin_beneddine_leclercq_garnier_bur_2021,dwivedi_sidharth_jovanovic_2022,Mauriello_Sharma_Larchevêque_Sandham_2025}.  

 Beyond modal instabilities, non-modal mechanisms can yield substantial amplification even when all eigenmodes are stable. \ac{RA} is particularly useful for identifying optimal forcing/response structures, i.e. disturbances of minimal input energy that generate large flow responses \cite{jovanovic2005componentwise,sipp2013characterization}. Streaky structures have been documented experimentally and numerically in \ac{SWBLI}s \cite{Tokura2011,sandham_schülein_wagner_willems_steelant_2014,Tong2017,sidharth2018,Dwivedi2019,Dwivedi2020,fang_zheltovodov_yao_moulinec_emerson_2020,lugrin_beneddine_leclercq_garnier_bur_2021,Duan2021compression_decompression,dwivedi_sidharth_jovanovic_2022,cao_hao_klioutchnikov_wen_olivier_heufer_2022,hao_2023}. Several mechanisms have been proposed for their generation. One is the lift-up effect, whereby streamwise vortices extract energy from the mean shear and produce transiently amplified, elongated low-/high-momentum streaks \cite{pirozzoli_bernardini_grasso_2010,Li2010DNScompressionramp,zuo_memmolo_huang_pirozzoli_2019,Dwivedi2019,lugrin_beneddine_leclercq_garnier_bur_2021,bugeat_robinet_chassaing_sagaut_2022}. Another involves centrifugal effects when the boundary layer experiences concave curvature, either due to geometry or due to curvature induced by displacement effects of the separation bubble; in this case, G\"ortler-like vortices may develop \cite{floryan_saric_1982,hall_malik_1989,floryan1991,schrader_brandt_zaki_2011,roghelia_olivier_egorov_chuvakhov_2017,Ren2018,shinde2019,currao_2020,li_choudhari_paredes_2022}. These vortices arise from an imbalance between destabilising centrifugal forces and viscous diffusion \cite{Gortler1941,Saric1994}, leading to spanwise modulation and streak formation. Other studies have emphasised streamwise deceleration near reattachment and baroclinic vorticity generation as additional contributors to streak formation \cite{Dwivedi2019,Dwivedi2020PRF,dwivedi_sidharth_jovanovic_2022}. In practice, these mechanisms need not be mutually exclusive; their relative importance depends on bubble strength and topology, geometry, and the upstream disturbance environment.  

\subsection{Non-linear mechanisms for laminar--turbulent transition}\label{subsec:non-linear mechanisms for laminar--turbulent transition}
Linear stability tools are essential for identifying and classifying primary instabilities, but by construction they describe infinitesimal disturbances and cannot capture the fully nonlinear evolution to breakdown. \Ac{WNL} analyses, based on asymptotic amplitude expansions, have provided insight into finite-amplitude evolution of primary instabilities and the emergence of secondary instabilities through nonlinear interactions. Notably, Dwivedi \emph{et al.}~\cite{dwivedi_sidharth_jovanovic_2022} showed that oblique waves developing over the separation bubble in a hypersonic double-wedge flow interact via quadratic nonlinearities to generate streamwise vortical excitations and streaks downstream of reattachment. Moreover, the production of low-frequency disturbances through nonlinear interactions among medium-frequency shear-layer fluctuations has been proposed as a potential contributor to the low-frequency unsteadiness observed in laminar \ac{SWBLI}s \cite{Sansica_Sandham_Hu_2016,mauriello2022,Mauriello_Sharma_Larchevêque_Sandham_2025,Saidi_Wang_Fournier_Tenaud_Robinet_2025}. Nevertheless, \ac{WNL} approaches are not designed to describe the full pathway to breakdown when dynamics become strongly nonlinear and involve multiple coupled mechanisms across multiple harmonics and scales. 

To overcome these limitations, fully nonlinear numerical approaches such as \ac{DNS} and \ac{LES} have been employed \cite{Sansica_Sandham_Hu_2016,Khotyanovsky2019,Shankar2019transitionalSWBLI,lugrin_beneddine_leclercq_garnier_bur_2021,sawant_theofilis_levin_2022,Mauriello_Sharma_Larchevêque_Sandham_2025,dixit_kumar_vadlamani_tsuboi_2025,caillaud2025}. These high-fidelity simulations with pre-defined forcing have enabled detailed characterisation of shock-induced transition, including saturation of primary instabilities, nonlinear generation of secondary instabilities (e.g. streamwise vortices and streaks), the emergence of coherent $\Lambda$-shaped structures, and the development of turbulent spots that ultimately merge into fully developed turbulence \cite{lugrin_beneddine_leclercq_garnier_bur_2021,cao_hao_klioutchnikov_olivier_wen_2021,cao_hao_klioutchnikov_wen_olivier_heufer_2022,dwivedi_sidharth_jovanovic_2022}. Complementary experiments using high-speed \ac{PIV}, pressure-sensor arrays, and Schlieren imaging have provided direct evidence of these evolving patterns, including signatures consistent with triadic interactions and secondary instabilities \cite{kontis_erdem_johnstone_murray_steelant_2013,roghelia_olivier_egorov_chuvakhov_2017,Giepman_Schrijer_vanOudheusden_2018,Threadgill2020,jiao_ma_xue_wang_chen_cheng_2024}. However, the computational cost of temporally and spatially resolved simulations of transitional \ac{SWBLI}s limits their use in systematic design and optimisation studies. For example, while \ac{DNS} can resolve the full range of scales involved in transition, it remains impractical for identifying optimal external disturbances---i.e. the minimal input or seed in space and time required to trigger breakdown---through extensive parametric exploration \cite{cherubini_depalma_robinet_bottaro_2011,pringle_willis_kerswell_2012}. 

\subsection{Contribution of this work}\label{subsec:contribution of this work}
Here we address these gaps by identifying the optimal nonlinear mechanisms governing convective-instability-driven transition in an oblique \ac{SWBLI} (in a regime without absolute/global instability). Our objective is to characterise the receptivity of the separated shear-layer/shock system to finite-amplitude external disturbances by determining the optimal nonlinear forcing in the frequency domain that leads to the most efficient pathway from such disturbances to breakdown. To this end, we consider a Mach 2.15 oblique-shock impingement configuration in which the shock-induced \ac{APG} is not sufficiently strong to yield a globally unstable laminar separation bubble. We build on nonlinear input--output analyses by Rigas \emph{et al.} \cite{rigas2021HBM} for incompressible boundary layers, Savarino \emph{et al.} \cite{Savarino_Sipp_Rigas_2025} for incompressible separated shear layers, and Poulain \emph{et al.} \cite{poulain2024} for compressible boundary layers, and extend these ideas to explicitly account for shocks and shock-induced separation within a Harmonic-Balanced Navier--Stokes (HBNS) / Space--Time Spectral Method (STSM) framework. The approach captures periodic, \ac{3D} nonlinearities by projecting the compressible \ac{N-S} equations into spectral space in time and homogeneous spatial directions and retaining a finite number of harmonics. Unlike linear analyses, it accounts for energy transfer among triads and thus enables us to describe a connected sequence of transitional events from the laminar separated state to breakdown. By formulating an adjoint-based nonlinear input--output optimisation, we seek the minimal external forcing of finite amplitude required to achieve the maximum increase in mean skin friction, thereby identifying the most efficient transition scenario. Beyond advancing the physical understanding of transitional \ac{SWBLI}s, this provides a computationally efficient route toward transition prediction and establishes a foundation for future flow-control and design-optimisation strategies in high-speed separated flows. 

The manuscript is organised as follows. In \S\ref{sec:theoretical background} we outline the methodology and numerical implementation for nonlinear input--output calculations. In \S\ref{sec:configuration and base-flow} we present the \ac{SWBLI} configuration and the laminar base flow. In \S\ref{sec:optimal non-linear mechanisms for laminar--turbulent transition} we discuss the optimal nonlinear mechanisms underpinning laminar--turbulent transition in the present \ac{SWBLI}. Conclusions and directions for future work are given in \S\ref{sec:conclusions}. Supplemental information on the base flow, linear analyses, additional parametric studies, a G\"ortler analysis, and a \ac{ZPG} boundary-layer benchmark is provided in Appendices \S\ref{appA}--\S\ref{appE}.

\section{\label{sec:theoretical background}Numerical methodology}

In this section we introduce the theoretical and numerical framework used to study laminar–turbulent transition in the \ac{SWBLI}. Specifically, we introduce governing equations, the non-linear input/output analysis implemented in the \ac{STSM} code and an overview of the numerical code.

\subsection{Governing equations}\label{subsec:governing equations}
We consider the compressible \ac{N-S} written in conservative form for the variables $(\rho, \rho \mathbf{u}, \rho E)$, 
\begin{subequations}
    \begin{equation}
        \frac{\partial \rho}{\partial t} + \nabla \cdot \left(\rho \mathbf{u} \right) = 0,
        \label{subeq:mass}
    \end{equation}
    \begin{equation}
        \frac{\partial (\rho \mathbf{u})}{\partial t} + \nabla \cdot \left(\rho \mathbf{u} \mathbf{u} + p\mathbf{I} - \boldsymbol{\tau} \right) = \mathbf{0},
        \label{subeq:momentum}
    \end{equation}
    \begin{equation}
         \frac{\partial (\rho E)}{\partial t} + \nabla \cdot \left[ (\rho E + p) \mathbf{u} - \boldsymbol{\tau} \cdot \mathbf{u} - \lambda \nabla T \right] = 0, 
        \label{subeq:energy}
    \end{equation}
    \label{eq:governing equations}
\end{subequations}
where $\rho$ is the density, $\mathbf{u}=[u,v,w]$ is the velocity vector with streamwise $u$, wall-normal $v$ and spanwise $w$ components, $p$ is the pressure, $E = p / (\rho (\gamma -1)) + \frac{1}{2}\mathbf{u} \cdot \mathbf{u}$ is the total energy, $\boldsymbol{\tau} = \mu (\nabla \mathbf{u} + (\nabla \mathbf{u})^\top) - \frac{2}{3} \mu (\nabla \cdot \mathbf{u}) \mathbf{I}$ is the viscous stress tensor, $\mathbf{I}$ the identity matrix, $T$ the static temperature, $\mu$ the viscosity, $\gamma=c_{p}/c_{v}$ the heat capacity ratio and $\lambda = \mu c_p/Pr$ where $c_p$ is the isobaric heat capacity and $Pr$ the Prandtl number. The system is closed with the  perfect-gas equation of state,
\begin{equation}
    p = \rho R T,
    \label{eq:ideal gas law}
\end{equation}
where $R$ is the specific gas constant, and Sutherland's law \cite{sutherland1893lii},
\begin{equation}
    \mu (T) = \mu_{\text{ref}} \left(\frac{T}{T_{\text{ref}}}\right)^{3/2} \frac{T_{\text{ref}} + S}{T + S},
    \label{eq:sutherland}
\end{equation}
with $S = 110.4$ $\text{K}$ being the Sutherland's temperature, $\mu_{\text{ref}} = 1.716 \times 10^{-5}$ $\text{kg}\text{m}^{-1}\text{s}^{-1}$ and $T_{\text{ref}} = 273.15$ $\text{K}$.

Introducing the conservative state $\mathbf{q}=[\rho, \rho \mathbf{u}, \rho E]^\top$, and considering an external momentum forcing term $\mathbf{f}=[f_{x},f_{y},f_{z}]^{\top}$ with amplitude $A$, eq.~\eqref{eq:governing equations} can be recast in compact state-space form,
\begin{equation}
    \frac{\partial \mathbf{q}}{\partial t} + \underbrace{\nabla \cdot \mathcal{F}(\mathbf{q})}_{\mathcal{N}(\mathbf{q})} = A\mathcal{P}\mathbf{f},
    \label{eq:state-space form}
\end{equation}
where $\mathcal{F}(\mathbf{q})$ contains the viscous and inviscid fluxes and $\mathcal{N}(\mathbf{q})$ denotes the non-linear differential \ac{N-S} operator. The $\mathcal{P}$ denotes a prolongation operation of the 3-state volumetric forcing $\mathbf{f}$ (acting only on momentum equations in this study, similar to the study of \citep{bugeat_robinet_chassaing_sagaut_2022}) to a 5-state vector to match the number of states on the left-hand-side of eq.~\eqref{eq:state-space form}.

\subsection{Non-linear input/output analysis}\label{subsec:non-linear input/output analysis}
\begin{figure}
    \centering
    \includegraphics[width=1\textwidth]{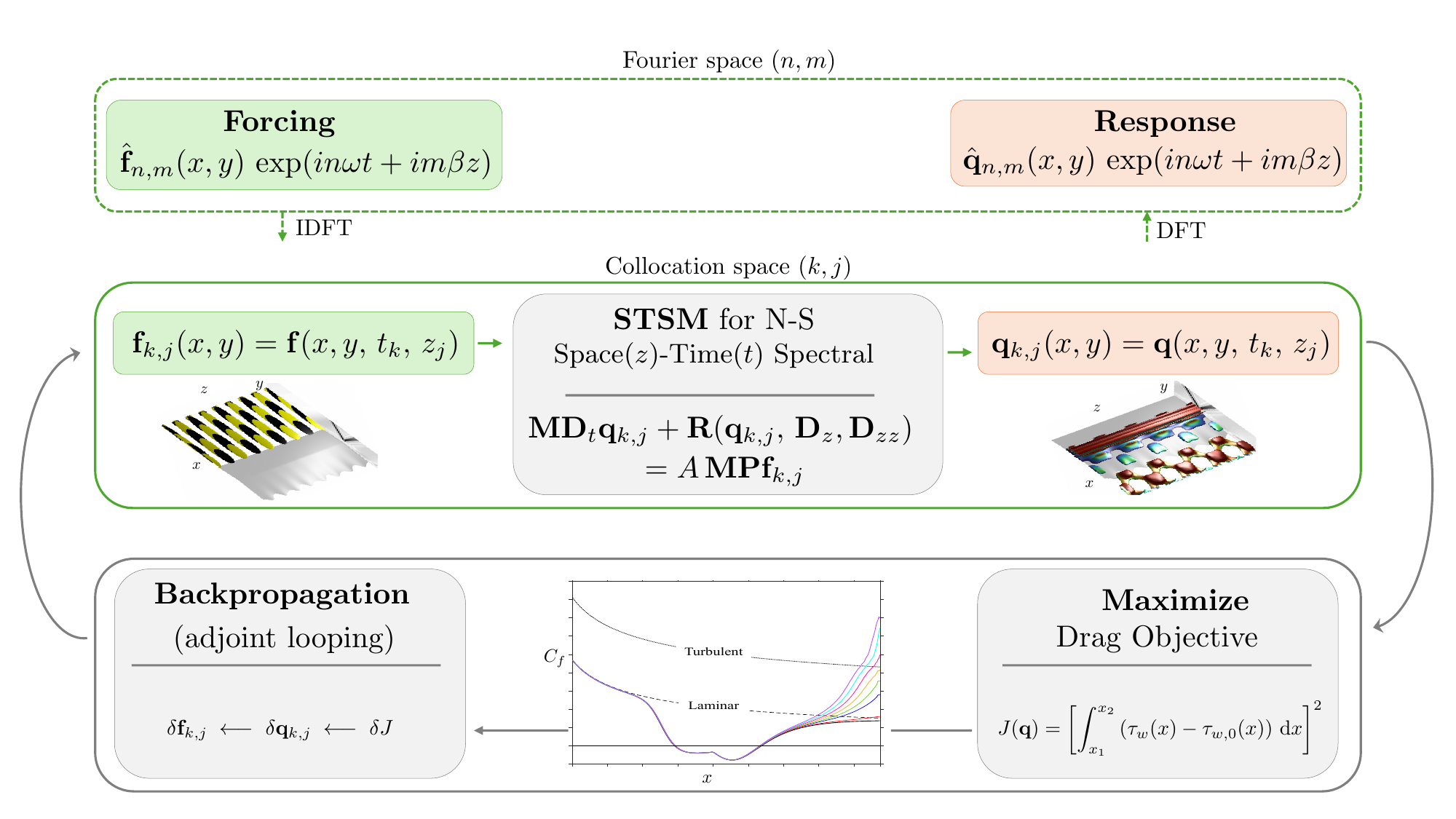}
\caption{ Schematic of the nonlinear input--output optimization framework based on the Space--Time Spectral Method (STSM). The forcing and response are represented by truncated Fourier modes \((n,m)\) and transformed by DFT/IDFT to the space--time collocation points \((k,j)\), where the nonlinear compressible Navier--Stokes residual is evaluated. For a prescribed forcing amplitude \(A\), the STSM system is solved by Newton iteration, the drag-based objective \(J(\mathbf q)\) is evaluated from the mean skin-friction increase, and its sensitivity is propagated backwards through the adjoint equations. The forcing is updated until convergence, yielding the optimal nonlinear forcing--response pair while avoiding explicit construction of harmonic interaction operators.
}
    \label{fig:Fig1}
\end{figure}

Non-linear input/output analysis is an extension of the linear frequency-domain resolvent framework, in which case a transfer function relating a single monochromatic forcing input state to its corresponding response output state is derived -refer to \S\ref{appB} for a brief review. This linear model is inadequate to characterize the complete transition process of fluid flow. The non-linear extension introduced by Rigas \emph{et al.} \cite{rigas2021HBM} for incompressible flows and adapted by Poulain \emph{et al.} \cite{poulain2024} for the compressible regime overcomes this limitation, shown schematically in Figure \ref{fig:Fig1}. Below, we review the fundamentals, although more details can be found in \cite{rigas2021HBM,poulain2024}. 

Upon discretization on a 2D $(x,y)$ grid, while keeping the spanwise spatial direction $z$ and time $t$ variables continuous, eq.~\eqref{eq:state-space form} can be rewritten in semi-discrete form,
\begin{equation}
    \mathbf{M} \frac{\partial \mathbf{q}}{\partial t}(z,t) + \mathbf{R} (\mathbf{q}) = A\mathbf{MP}\mathbf{f}(z,t),
    \label{eq:semi-discrete N-S}
\end{equation}
where $\mathbf{M}$ is the mass matrix associated with the $(x,y)$ discretization, $\mathbf{P}$ is the discrete prolongation matrix and $\mathbf{R} (\mathbf{q})$ is the non-linear residual of the compressible \ac{N-S}.

In the present wall-bounded configuration we assume periodicity in the spanwise direction and consider disturbances that are statistically homogeneous in $z$ over a chosen spanwise period $L_z$. Moreover, we consider  time-periodic forcing at a prescribed fundamental frequency (or to a finite set of discrete forcing frequencies). For a deterministic periodic input, the long-time response of the nonlinear \ac{N-S} equations can be sought in the form of a periodic solution, for which a Fourier series in time is a natural representation. Nonlinear terms then generate harmonics and cross-interactions; retaining a finite number of harmonics provides a controlled approximation that captures mean-flow distortion and the dominant triadic energy transfers while remaining computationally tractable. Accordingly, we expand both the forcing and the flow state in a truncated Fourier basis in $(t,z)$,
\begin{subequations}
    \begin{align}
        \mathbf{f}(z,t) &= \sum_{\substack{n=-N,m=-M, \\ (n,m) \neq (0,0)}}^{N,M} \hat{\mathbf{f}}_{n,m}(n\omega,m\beta;x,y) \exp\left [ \mathrm{i}\left ( n \omega t + m \beta z \right ) \right ], \\[6pt]    
        \mathbf{q}(z,t) = \hat{\mathbf{q}}_{0,0}(x,y) 
        &+ \sum_{\substack{n=-N,m=-M, \\ (n,m) \neq (0,0)}}^{N,M} \hat{\mathbf{q}}_{n,m}(n\omega,m\beta;x,y) \exp\left [ \mathrm{i}\left ( n \omega t + m \beta z \right ) \right ],
    \end{align}
    \label{eq:Fourier expansion HBNS}
\end{subequations}
\noindent where the symbol ${\hat{(\cdot)}}$ denotes the complex Fourier coefficients, $\omega$ and $\beta$ are the fundamental temporal frequency and spanwise wavenumber, and $N,M$ denote the maximum number of harmonics retained. The zero-th harmonic $\hat{\mathbf{q}}_{0,0}$ is the time- and spanwise-averaged state (hereafter referred to as the mean flow). By construction, we set the zero-th forcing harmonic to zero (i.e. no steady forcing allowed). 
Since the physical state $\mathbf{q}$ is real, the Fourier coefficients satisfy the symmetry $\hat{\mathbf{q}}_{-n,-m}=\hat{\mathbf{q}}^{*}_{n,m} $ for all $(n,m)$, which also  implies that $\hat{\mathbf{q}}_{0,0}$ is real. The superscript $^{*}$ denotes complex conjugation. The same symmetry applies to the forcing.

Substituting the expansions~\eqref{eq:Fourier expansion HBNS} into~\eqref{eq:semi-discrete N-S} and collecting terms at each $(n,m)$ yields the non-linear \ac{HBNS} system,
\begin{equation} 
    \mathrm{i}n\omega\mathbf{M}\hat{\mathbf{q}}_{n,m} + \hat{\mathbf{R}}_{n,m}(\mathbf{q}) = A \mathbf{M}\mathbf{P}\hat{\mathbf{f}}_{n,m},
    \label{eq:HBNS compact}
\end{equation}
\noindent where
\begin{equation}
    \hat{\mathbf{R}}_{n,m}(\mathbf{q}) = \hat{\mathbf{L}}(\hat{\mathbf{q}}_{n,m}) + \sum_{\substack{-N<a,b,c<N \\ -M<d,e,f<M \\ a+b+c=n \\ d+e+f=m}}\hat{\mathbf{N}} (\hat{\mathbf{q}}_{a,d},\hat{\mathbf{q}}_{b,e},\hat{\mathbf{q}}_{c,f})
    \label{eq:HBNS residual}
\end{equation}
\noindent is the discrete residual in frequency domain containing linear $\hat{\mathbf{L}}$ and non-linear $\hat{\mathbf{N}}$ terms. 
The equation in \eqref{eq:HBNS compact} obtained for $n=m=0$ governs the evolution of the mean-flow (as it departs from the base-flow due to the nonlinear Reynolds stresses of the fluctuating harmonics); the subsequent equations govern the evolution of each $(n,m)$ perturbation mode.

For incompressible flows with quadratic nonlinearity, system~\eqref{eq:HBNS compact} can be handled analytically in the frequency domain using the \ac{AHBM} \cite{rigas2021HBM,Savarino_Sipp_Rigas_2025}, by constructing the corresponding triadic convolution operators explicitly. In compressible flows, however, the nonlinear terms are not purely quadratic and, in particular, variable transport properties introduce non-polynomial dependencies (e.g. through Sutherland’s law). As a result, an explicit construction of all nonlinear convolution terms in \eqref{eq:HBNS residual} becomes prohibitively expensive as the number of retained harmonics increases, especially for three-dimensional spanwise-periodic disturbances where many $(\omega,\beta)$ combinations must be coupled.

To date, Sierra-Ausin \emph{et al.} \cite{sierra_citro_giannetti_fabre_2022} have employed the \ac{AHBM} for compressible flows for \ac{2D} unsteady disturbances. To circumvent the computational bottleneck associated with explicit convolution construction in the present setting, we instead resort to the \ac{STSM}, as implemented in Poulain \emph{et al.} \cite{poulain2024}  and described in the next subsection, which enforces the same HBNS truncation but evaluates the nonlinear terms in a pseudo-spectral time--spanwise collocation framework (transforming between physical and spectral space), thereby avoiding the explicit assembly of high-dimensional convolution operators.

\subsection{Space-Time Spectral Method}\label{subsec:STSM}
The \ac{STSM} is a pseudo-spectral collocation method \cite{gopinath2005time,liu2006comparison} that enforces the truncated Fourier representation of the HBNS system using a discrete set of collocation points in time and spanwise direction. Instead of solving the frequency-domain \ac{HBNS} system~\eqref{eq:HBNS compact} directly for the $t$-harmonics $n\in[-N,-N+1,\dots,-1,0,1,\dots,N]$ and the $z$-harmonics $m\in[-M,-M+1,\dots,-1,0,1,\dots,M]$, we introduce $(2N+1)$ equispaced collocation points $k=[0,1,\dots,2N]$ in time and $(2M+1)$ equispaced collocation points $j=[0,1,\dots,2M]$ in span,
\begin{equation}
    \begin{aligned}
    &\begin{array}{cccccc}
    z \backslash t & -N & \cdots & n & \cdots & N \\
    -M\ & \ddots & \cdots & \cdots & \cdots & \ddots \\
    \vdots & \vdots & \ddots & \ddots & \ddots & \vdots \\
    m  & \vdots & \ddots & \hat{\mathbf{q}}_{n,m}\exp{\left[\mathrm{i}(n\omega t+ m\beta z)\right]}  & \ddots & \vdots \\
    \vdots & \vdots & \ddots & \ddots & \ddots & \vdots \\
    M & \ddots  & \cdots & \cdots & \cdots & \ddots \\
    \end{array} \\[6pt]
    &\Longrightarrow \quad
    \begin{array}{cccccc}
    z \backslash t & 0 & \cdots & k & \cdots & 2N \\
    0 & \ddots & \cdots & \cdots  & \cdots & \ddots \\
    \vdots & \vdots & \ddots & \ddots & \ddots & \vdots \\
    j  & \vdots  & \ddots & \mathbf{q}_{k,j}=\mathbf{q}(t_k=k\Delta t,z_j=j\Delta z)  & \ddots & \vdots \\
    \vdots & \vdots & \ddots & \ddots & \ddots & \vdots \\
    2M  & \ddots  & \cdots & \cdots & \cdots & \ddots \\
    \end{array}
    \end{aligned}
    \label{eq:frequency-to-collocation-points}
\end{equation}
\noindent where $\Delta t=(2\pi/\omega)/(2N+1)$ and $\Delta z=(2\pi/\beta)/(2M+1)$ (with spanwise period $L_z=2\pi/\beta$). For a $(t,z)$-periodic solution truncated to $|n|\le N$ and $|m|\le M$, the Fourier coefficients $\hat{\mathbf{q}}_{n,m}$ and the collocation values $\mathbf{q}_{k,j}$ are equivalent representations: $(2N+1)(2M+1)$ collocation values uniquely determine the retained Fourier coefficients, and vice versa.

Enforcing the governing equations at each collocation node yields the collocation form
\begin{equation}
    \mathbf{M}\,\frac{\partial \mathbf{q}_{k,j}}{\partial t} + \mathbf{R}(\mathbf{q}_{k,j}) = A\,\mathbf{MP}\,\mathbf{f}_{k,j},
    \label{eq:STSM}
\end{equation}
\noindent where $\mathbf{R}(\cdot)$ denotes the nonlinear residual operator. Unlike the fully analytic harmonic-balance formulation, system \eqref{eq:STSM} does not require explicit construction of convolution operators for nonlinear mode coupling. This is particularly advantageous in compressible flows, where nonlinearities are not purely quadratic and where temperature-dependent transport properties (e.g. Sutherland's law) introduce non-polynomial dependencies, making the explicit assembly of mode-coupling operators prohibitively expensive for three-dimensional spanwise-periodic disturbances.

The $t$- and $z$-derivatives appearing in $\partial \mathbf{q}_{k,j}/\partial t$ and $\mathbf{R}(\mathbf{q}_{k,j})$ are evaluated pseudo-spectrally using precomputed differentiation matrices,
\begin{equation}
    \frac{\partial \mathbf{q}_{k,j}}{\partial t} = \sum_{k'=0}^{2N} (\mathbf{D}_t)_{kk'}\,\mathbf{q}_{k',j},\;\;\;\;\;\;\;\; 
    \frac{\partial \mathbf{q}_{k,j}}{\partial z} = \sum_{j'=0}^{2M} (\mathbf{D}_z)_{jj'}\,\mathbf{q}_{k,j'},\;\;\;\;\;\;\;\; 
    \frac{\partial^2 \mathbf{q}_{k,j}}{\partial z^2} = \sum_{j'=0}^{2M} (\mathbf{D}_{zz})_{jj'}\,\mathbf{q}_{k,j'} ,
    \label{eq:derivatives t z zz}
\end{equation}
\noindent constructed via discrete Fourier transforms. For the time direction, we define the \ac{DFT}
\begin{equation}
    (\mathbf{E}_t)_{nk} = \frac{1}{2N+1}\exp\left[-\frac{\mathrm{i}2\pi kn}{2N+1}\right]\mathbf{I},
    \label{eq:DFT}
\end{equation}
\noindent and the \ac{IDFT}
\begin{equation}
    (\mathbf{E}_t^{-1})_{kn} = \exp\left[\frac{\mathrm{i}2\pi kn}{2N+1}\right]\mathbf{I},
    \label{eq:IDFT}
\end{equation}
\noindent such that $\hat{\mathbf{q}}_{n,j}=\sum_{k=0}^{2N}(\mathbf{E}_t)_{nk}\mathbf{q}_{k,j}$. The time derivative then reads
\begin{equation}
    \frac{\partial \mathbf{q}_{k,j}}{\partial t}
    = \sum_{k'=0}^{2N} (\mathbf{D}_t)_{kk'}\,\mathbf{q}_{k',j}
    = \sum_{n=-N}^{N}\sum_{k'=0}^{2N}(\mathbf{E}_t^{-1})_{kn} (\mathrm{i} n \omega \mathbf{I}) (\mathbf{E}_t)_{nk'} \mathbf{q}_{k',j},
    \label{eq:time derivative}
\end{equation}
\noindent where $(\mathrm{i} n \omega \mathbf{I})$ is block diagonal, with blocks equal to the identity matrix scaled by $\mathrm{i}n\omega$. Analogous constructions are used for $\mathbf{D}_z$ and $\mathbf{D}_{zz}$ (not shown for brevity). Since the matrices $\mathbf{D}_{(\cdot)}$ depend only on $(N,M)$, they are assembled once and reused during the residual evaluation.

Lastly, the discretised collocation STSM system can be written as
\begin{equation}
    \mathbf{M} \mathbf{D}_t \mathbf{q}_{k,j}
    + \mathbf{R}_\text{2-D} (\mathbf{q}_{k,j})
    + \mathbf{R}_z \left(\mathbf{q}_{k,j},\mathbf{D}_z\mathbf{q}_{k,j}\right)
    + \mathbf{R}_{zz} \left(\mathbf{q}_{k,j},\mathbf{D}_{zz}\mathbf{q}_{k,j}\right)
    = A\mathbf{MP}\mathbf{f}_{k,j},
    \label{eq:full residual}
\end{equation}
\noindent and is solved using a Newton algorithm (details in \cite{poulain2024}).

\subsubsection{Optimal forcing for maximal drag increase}\label{subsubsec:optimal forcing for maximal drag increase}
The forcing input is generally not known \emph{a priori}, except in case-specific control studies where the forcing shape is constrained by actuator capabilities or by prior physical insight \cite{Michelis2017impulsiveforcing,Yarusevych2017steady,Michelis2018spanwise}. We therefore augment the nonlinear STSM solver with an adjoint-based optimisation procedure that seeks an ``optimal'' forcing structure $\mathbf{f}_{k,j}$ that maximises a prescribed indicator of transition (here a drag proxy).

In classic linear resolvent analysis, the objective is often the energy gain, i.e. the ratio between the energy of the monochromatic response and that of the forcing  (see Appendix~\ref{appB}). In the present nonlinear framework, instead, we directly target a drag-related quantity based on the mean wall shear stress \cite{rigas2021HBM,poulain2024}. Denoting by $\overline{(\cdot)}$ the time average over one forcing period, we define
\begin{equation}
    J(\mathbf{q}) = \left[ \int_{x_1}^{x_2} \left(\overline{\tau}_w(x) - \tau_{w,0}(x)\right)\, \mathrm{d}x \right]^2,
    \qquad
    \tau_w(x)=\mu\,\left.\frac{\partial u}{\partial y}\right|_{y=0},
    \label{eq:cost function}
\end{equation}
where $\tau_{w,0}$ is the wall shear stress of the laminar base flow $\mathbf{q}_0$ and $\mu$ is the dynamic viscosity. The integral in \eqref{eq:cost function} is proportional to the skin-friction drag on the plate; thus maximising $J$ promotes the largest mean drag increase associated with transition onset. The square ensures a non-negative objective and improves numerical scaling; in the regimes of interest here, $\int (\overline{\tau}_w-\tau_{w,0})\,dx$ is observed to be positive, so maximising $J$ is equivalent to maximising the drag increase itself.

The optimisation is posed as a constrained variational problem. The forcing amplitude is prescribed using a discrete $L^2$ norm,
\begin{equation}
    \|\mathbf{f}\|_F^2 \equiv \mathbf{f}^* \mathbf{Q}_F \mathbf{f} = A^2,
    \label{eq:forcing_norm}
\end{equation}
where $\mathbf{Q}_F$ is the Hermitian positive-definite matrix defining the discrete $L^2$ inner product in forcing space; in the present finite-volume discretisation, $\mathbf{Q}_F$ reduces to a diagonal quadrature-weight matrix containing the cell areas/volumes. The state must satisfy the nonlinear governing equations~\eqref{eq:semi-discrete N-S}. Introducing the adjoint variable $\tilde{\mathbf{q}}$ and the scalar multiplier $\lambda$, we define the Lagrangian
\begin{equation}
    \mathcal{L} (\mathbf{f},\mathbf{q},\tilde{\mathbf{q}},\lambda)
    = J(\mathbf{q})
    - \left\langle \tilde{\mathbf{q}},\, \mathbf{M}\frac{\partial \mathbf{q}}{\partial t}+\mathbf{R}(\mathbf{q})-\mathbf{M}\mathbf{P}\mathbf{f}\right\rangle
    - \left\langle \lambda,\, \mathbf{f}^* \mathbf{Q}_F \mathbf{f} - A^2 \right\rangle,
    \label{eq:Lagrangian}
\end{equation}
where $\langle\cdot,\cdot\rangle$ denotes the discrete inner product consistent with the spatial discretisation.

Setting the first variations of $\mathcal{L}$ to zero yields the adjoint system
\begin{equation}
    \underbrace{\left( \mathbf{M}\frac{\partial}{\partial t} + \frac{\partial \mathbf{R}(\mathbf{q})}{\partial \mathbf{q}} \right)^{\dagger}}_{\mathbf{J}^{\dagger}}
    \tilde{\mathbf{q}}
    = \frac{\mathrm{d} J(\mathbf{q})}{\mathrm{d} \mathbf{q}},
    \label{eq:adjoint system}
\end{equation}
together with the optimality condition obtained by variation with respect to $\mathbf{f}$,
\begin{equation}
    2\lambda\,\mathbf{f}
    = \underbrace{\mathbf{Q}_F^{-1}\mathbf{P}^*\mathbf{M}^*\,\tilde{\mathbf{q}}}_{\check{\mathbf{q}}}
    \equiv \check{\mathbf{q}},
    \label{eq:parallelism condition}
\end{equation}
which shows that the optimal forcing is parallel to the restricted adjoint state $\check{\mathbf{q}}$ (up to the scalar factor $1/(2\lambda)$). Accordingly,
\begin{equation}
    \cos\theta
    = \frac{\mathbf{f}^* \mathbf{Q}_F \check{\mathbf{q}}}{A\,\gamma}=1,
    \qquad
    \gamma^2=\check{\mathbf{q}}^*\mathbf{Q}_F\check{\mathbf{q}},
    \label{eq:cos theta}
\end{equation}
and the angle $\theta$ provides a convenient convergence criterion for the iterative optimisation.

\begin{figure}
    \centering
    \includegraphics[width=0.81\textwidth]{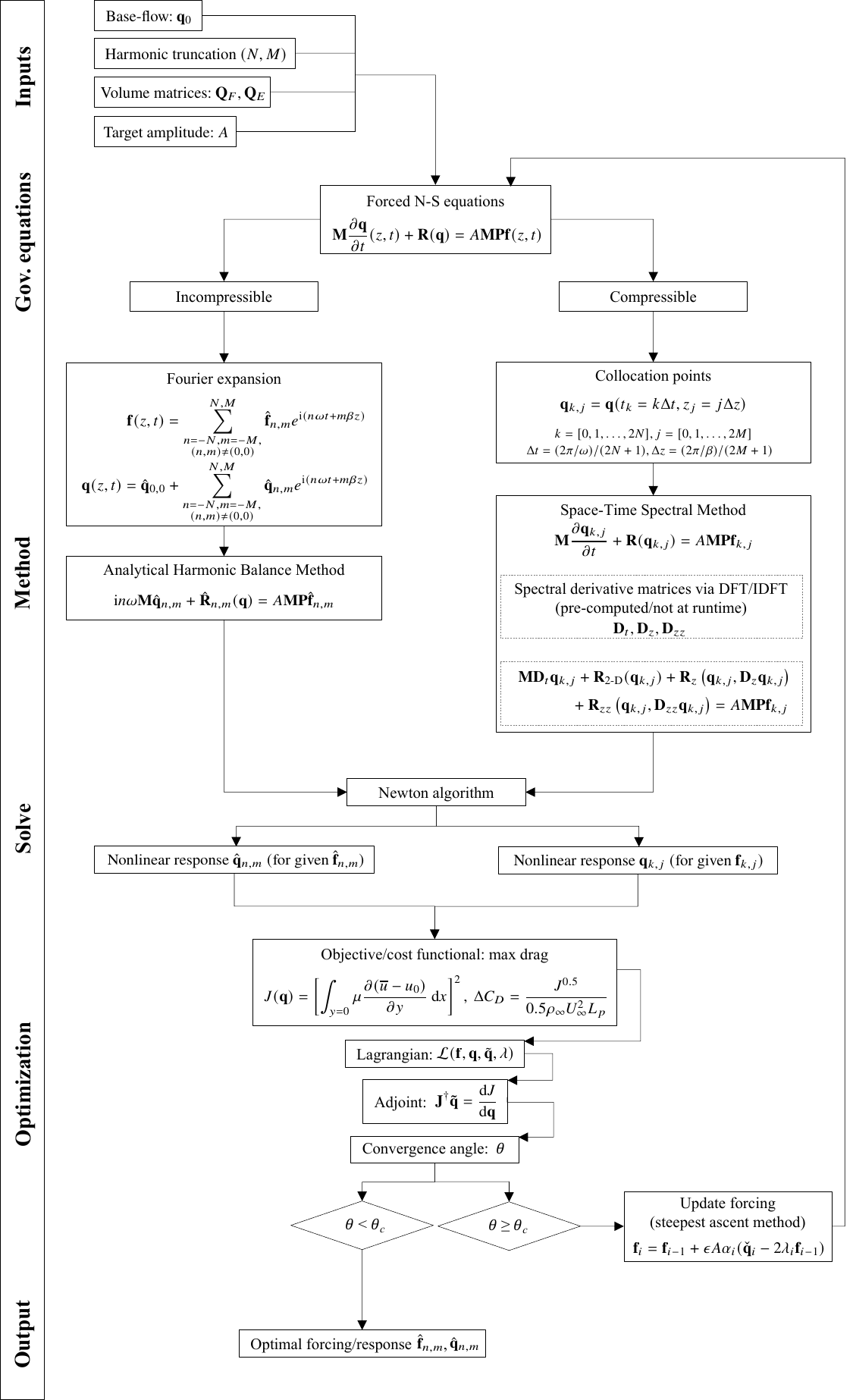}
    \caption{Detailed schematic of the nonlinear frequency-domain input--output optimisation framework for \ac{SWBLI} transition.}
    \label{fig:framework}
\end{figure}

An outline of the optimisation procedure is given in Algorithm~\ref{alg:optimization}, and a compact illustration of the complete methodology is shown in Fig.~\ref{fig:framework}.

\begin{algorithm}
\label{alg:optimization}
\textbf{Optimisation algorithm to compute the optimal nonlinear forcing.}

\begin{algorithmic}[1]
\State Set the step-length parameter $c\in(0,1]$ (largest/smallest step length: $c\to 1$ / $c\to 0$)
\State Set $\epsilon \gets 1$ for maximisation, or $\epsilon \gets -1$ for minimisation
\State Set the tolerance $\theta_{c}$ for convergence of the angle $\theta$
\State Initialise the forcing $\mathbf{f}_0$ such that $\sqrt{\mathbf{f}_0^* \mathbf{Q}_F \mathbf{f}_0}\approx A$
\State Scale $\mathbf{f}_0$ to the prescribed amplitude $A$
\State Solve for the initial state from \eqref{eq:STSM} using a Newton method
\State $i\gets 0$
\Repeat
  \Comment{Adjoint loop for forcing optimisation}
  \State $i\gets i+1$
  \State Solve for the updated state $\mathbf{q}_i=\mathbf{q}(\mathbf{f}_{i-1})$ from \eqref{eq:STSM} (Newton)
  \State Solve for the adjoint state $\tilde{\mathbf{q}}_i$ from \eqref{eq:adjoint system}
  \State Compute the restricted adjoint state $\check{\mathbf{q}}_{i}$ from \eqref{eq:parallelism condition}
  \State Compute the convergence angle:
  $\theta_i = \arccos\!\left(\dfrac{\mathbf{f}_{i-1}^{*}\mathbf{Q}_F\check{\mathbf{q}}_{i}}{A \gamma_{i}}\right)$,
  where $\gamma_i^2 = \check{\mathbf{q}}_{i}^{*} \mathbf{Q}_F \check{\mathbf{q}}_i$
  \State Compute $\alpha_i=c/\gamma_i$ and the step length:
  $\lambda_i = \dfrac{\epsilon + c \cos \theta_i - \epsilon \sqrt{1-c^2\sin^2 \theta_i}}{2 A \alpha_i}$
  \State Update the forcing:
  $\mathbf{f}_{i} = \mathbf{f}_{i-1} + \epsilon A \alpha_i (\check{\mathbf{q}}_{i} - 2 \lambda_i \mathbf{f}_{i-1})$
  \State Scale $\mathbf{f}_{i}$ to the prescribed amplitude $A$
\Until{$\theta_i < \theta_c$}
\end{algorithmic}
\end{algorithm}

\subsection{Numerical code}\label{subsec:numerical_code}
We use two main codes: the open-source \textsc{BROADCAST} package \cite{poulain2022} for base-flow and linear stability calculations, and \ac{Pyst}, an in-house Python module implementing the \ac{STSM}. The \ac{STSM} extends the original 2D \ac{TSM} \cite{moulin2020flutter} to three-dimensional, spanwise-periodic disturbances. The \ac{STSM} code was validated on a supersonic (Mach 4.5) flat-plate boundary layer \cite{poulain2024} and is employed here for the \ac{SWBLI} configuration.

\subsubsection{BROADCAST}\label{subsubsec:BROADCAST}
The laminar base flow is computed using a fifth-order \ac{FE MUSCL} shock-capturing scheme. Viscous fluxes are discretized with a five-point compact stencil that is fourth-order accurate. \textsc{BROADCAST} also obtains exact derivatives of the compressible linearized \ac{N-S} operators via \ac{AD} using \textsc{TAPENADE} \cite{hascoet2013tapenade}, enabling linear global and resolvent analyses for supersonic flows with shocks. Depending on the computational cost, the linearized operators are applied either through exact LU factorization or via approximate \ac{GMRES} computations. Exploiting operator sparsity, linear systems are solved through the PETSc interface \cite{balay2019petsc}. For global stability and linear resolvent analyses, we use the SLEPc library \cite{roman2015slepc} (Krylov--Schur methods \cite{hernandez2007krylov}), and in particular the Arnoldi algorithm. Newton iterations are performed using a pseudo-transient relaxation strategy \cite{crivellini2011implicit} to aid convergence. The Newton correction is obtained with a \ac{GMRES} solver preconditioned by a block-circulant preconditioner, typically converging in a few iterations.

At present, \textsc{BROADCAST} handles 2D curvilinear, multiblock, structured meshes and the main flow solver runs sequentially. The associated linear algebra operations are distributed across multi-core architectures via PETSc/OpenMP/MPI \cite{lange_achieving_2013}.

\subsubsection{Pyst}\label{subsubsec:pyst}
\ac{Pyst} is interfaced with \textsc{BROADCAST} to perform the fundamental algorithmic operations of the \ac{STSM}. Because the \ac{STSM} is pseudo-spectral, aliasing arises from the nonlinearity of the residual; therefore, de-aliasing requires computing twice as many harmonics \cite{labryer2009high}. In the present application, the spanwise harmonics carry significantly more energy than the temporal harmonics, consistent with observations in incompressible separated shear-layer flows \cite{savarino2022laminar,Savarino_Sipp_Rigas_2025}. We therefore apply de-aliasing only in the spanwise direction. The only user inputs to \ac{Pyst} are the step-length parameter (between 0 and 1) and the tolerance for the convergence criterion (eq.~\eqref{eq:cos theta}), set here to $1^\circ$. \ac{Pyst} is parallelized with MPI: each MPI rank handles one temporal or spanwise instance of the system in eq.~\eqref{eq:STSM}. 

\section{\label{sec:configuration and base-flow}Configuration and base-flow}

The \ac{SWBLI} configuration is illustrated in this section, including the geometry, computational mesh, reference length scales and boundary conditions.

Following Robinet \cite{robinet2007bifurcations}, the reference length, velocity and time scales are $l_{\mathrm{ref}}=X_{sh}$, $U_{\mathrm{ref}}=U_{\infty}$ and $t_{\mathrm{ref}}=X_{sh}/U_{\infty}$, respectively, where $X_{sh}=0.08 \mathrm{~m}$ is defined as the distance between the inviscid shock impingement point (where $x/X_{sh}=1$) and the flat plate leading edge (where $x/X_{sh}=0$). The $\infty$ symbol denotes free-stream conditions. Streamwise, wall-normal and spanwise coordinates $(x,y,z)$ are therefore non-dimensionalized by $X_{sh}$. The reference Reynolds number is $Re=\rho_{\infty}U_{\infty}X_{sh}/\mu_{\infty}=10^{5}$.

\begin{figure}
    \includegraphics[width=\textwidth]{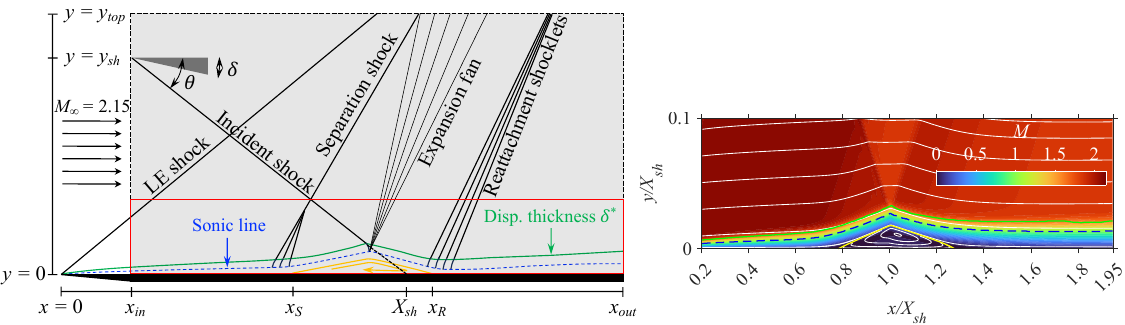}
    \caption{Left: schematic of the oblique \ac{SWBLI} problem configuration. The computational domain is marked in gray. The incident shock propagates from height $y_{sh}$ at angle $\theta$ relative to the streamwise direction. Right: zoomed-in view of the shock-induced separation of the laminar base-flow computed for $\theta=30.8^{\circ}$. The domain marked on the left by the red box is displayed. Contours of Mach number and streamlines are superimposed. The sonic and boundary layer displacement thickness lines are plotted with blue-dashed and green-solid lines. The dividing streamline of the separation bubble is plotted in yellow.}
    \label{fig:configuration}
\end{figure}

\begin{table}
    \caption{Domain set-up and reference scales for the $\theta=30.8^{\circ}$ \ac{SWBLI} base-flow calculation. The spatial coordinates are in non-dimensional units.}
    \centering
    \begin{tabular}{ccccccccccc}
    \hline
    $L_p \: (\mathrm{m})$ & $X_{sh} \: (\mathrm{m})$ & $U_{\mathrm{ref}} \: (\mathrm{m/s})$ & $t_{\mathrm{ref}} \: (\mathrm{10^{-3}s})$ & $x_{\mathrm{in}}$ & $x_{\mathrm{out}}$ & $y_{\mathrm{top}}$ & $\delta \: (^{\circ})$ & $\theta \: (^{\circ})$ & $Re_{\delta^{*}}$ & $Re_{L_{\mathrm{sep}}}$\\\hline
    0.156 & 0.08 & 540.1 & 0.15 & 0.20 & 1.95 & 0.60 & 3.81 & 30.8 & $2.7\times10^{3}$ & $5.1\times10^{4}$ \\
    \hline
    \end{tabular}
    \label{tab:domain set-up}
\end{table}

Figure \ref{fig:configuration} (left) shows schematically the geometry of the problem. A smooth flat plate of length $L_p=0.156\mathrm{~m}$ is considered. The computational domain is rectangular. Inlet and outlet are located at $x_{\mathrm{in}}=0.2$ and $x_{\mathrm{out}}=1.95$ referenced to the flat plate leading edge, while the top boundary is at $y_{\mathrm{top}}=0.60$. An oblique shock is generated from a (virtual) shock generator placed at $y/X_{sh}=0.48$ above the flat plate. By setting the origin and angle $\theta$ of the incident shock, we obtain a \ac{SWBLI} with a Mach 2.15 boundary layer. While $\theta$ is a parameter that can be easily adjusted in the set-up, we focus on the $\theta=30.8^{\circ}$ (correspondingly, $\delta=3.81^{\circ}$) case for the non-linear analysis. At the impingement point $Re_{\delta^{*}}=2.7\times10^3$, where $Re_{\delta^{*}}$ is the local Reynolds number based on the boundary layer displacement thickness. These parameters are summarized in table \ref{tab:domain set-up}.

\begin{table}
    \caption{Free-stream conditions and working gas properties.}
    \centering
    \begin{tabular}{ccccccc}
    \hline
    $M_{\infty}$ & $p_{\infty} \: (\mathrm{MPa})$ & $T_{0_{\infty}} \: (\mathrm{K})$ & $\rho_{\infty} \: (\mathrm{kgm^{-3}})$ & $\mu_{\infty} \: (\mathrm{kgm^{-1}s^{-1}})$ & $\gamma$ & $R \: (\mathrm{Jkg^{-1}K^{-1}})$ \\\hline
    2.15 & 0.0112 & 302 & 0.0248 & $1.0726\times10^{-5}$ & 1.4 & 287.1 \\
    \hline
    \end{tabular}
    \label{tab:free stream conditions}
\end{table}

The computational mesh is constructed with an equi-spaced distribution of $N_{x}$ points in the streamwise direction and with a bi-geometric distribution of $N_{y}$ points in the wall-normal direction. The number of grid points is varied to obtain grid convergence of the numerical base-flow solution. The working fluid is air modeled as an ideal gas with $\gamma=1.4$, $R=287.1\mathrm{~Jkg^{-1}K^{-1}}$ and $Pr=0.72$. At the inlet we impose a \ac{ZPG} boundary layer profile and shock jump conditions accounting for the presence of a weak shock emanating from the flat plate leading edge. Upstream of the leading edge shock the flow states are at free-stream conditions, which are outlined in table \ref{tab:free stream conditions}. We also impose a supersonic extrapolated outlet and non-reflective far field at the top boundary. Finally, we apply the no-slip and adiabatic wall boundary conditions on the plate. Pressure, temperature, density and velocities are referenced to the free-stream static quantities.

A close-up view of the shock-induced separated zone is displayed in figure \ref{fig:configuration} (right) for $\theta=30.8^{\circ}$. Colored contours of Mach number and streamlines from the converged base-flow highlight the key topological features of the flow, namely the incident and separation shocks, the expansion fan and the reattachment shocklets. Below the sonic line, a laminar separation bubble with characteristic length $Re_{L_{\mathrm{sep}}}=5.1\times10^{4}$ hosts recirculating flow. The extent of the separated flow is demarcated by the separation, where $Re_{x_{S}} = 0.78\times10^5$, and the reattachment, $Re_{x_{R}} = 1.29\times10^5$, points. The validation of the numerical base-flow is in \S\ref{appA}.

The $\theta=30.8^{\circ}$ configuration examined in this work is a convectively unstable (globally stable, see \S\ref{appB} for a detailed global linear stability and resolvent analysis) separated \ac{SWBLI}, whose transitional dynamics is governed by the non-linearities of the flow which are activated when the infinitesimal free-stream boundary layer disturbances reach sufficient energy (or amplitude) in the separated shear layer \cite{sandham_schülein_wagner_willems_steelant_2014,mauriello2022,Mauriello_Sharma_Larchevêque_Sandham_2025}. In agreement with existing literature \cite{bugeat_robinet_chassaing_sagaut_2022,hao_2023,zhao_ma_chen_zhang_hao_wen_2024}, we identify through the study of the linear resolvent operator three frequency-spanwise wavenumber regions where distinct mechanisms are active (see \S\ref{appB} for more detailed analysis). Primarily, the modal oblique first Mack wave instability typically found in compressible boundary layers \cite{Franko_Lele_2013,kuehl_paredes_2016,chen_chen_yuan_tu_zhang_2019,li_choudhari_paredes_2022,dwivedi_sidharth_jovanovic_2022,poulain2024,Mauriello_Sharma_Larchevêque_Sandham_2025} is the most linearly unstable mode at the Strouhal frequency based on the shock impingement distance $fX_{sh}/U_{\infty}=2$ and non-dimensional spanwise wavenumber $\beta X_{sh}=45$. At nominally zero frequency and $\beta X_{sh}=163$, streaks experience transient growth via the non-modal lift-up mechanism observed in both incompressible and compressible boundary layers (with and without flow separation) \cite{andersson_brandt_bottaro_henningson_2001,Dwivedi2019,Dwivedi2020PRF,rigas2021HBM,lugrin_beneddine_leclercq_garnier_bur_2021,bugeat_robinet_chassaing_sagaut_2022,poulain2024,zhao_ma_chen_zhang_hao_wen_2024,Jaroslawski_Forte_Vermeersch_Moschetta_Gowree_2024,Savarino_Sipp_Rigas_2025}. Finally, a third region of mild amplification hosts the modal bubble breathing mechanism, which is associated to a stable, zero-frequency, quasi-2D eigenvalue of the global linearized Jacobian operator \cite{robinet2007bifurcations,Hildebrand2018,cao_hao_klioutchnikov_wen_olivier_heufer_2022,hao_2023,song2023}. While this intrinsic mechanism predominantly drives the non-linear dynamics of large shock-induced separation bubbles \cite{song2023}, it is significantly damped compared to the other mechanisms in the present configuration.

The rich environment of various disturbances linear stability analyses reveal sets the ground for the study of the non-linear evolution of such disturbances towards the physical understanding of transition to turbulence in \ac{SWBLI}. Saturation of primary instabilities, energy transfer mechanisms and mean-flow modifications must be considered by accounting for non-linear self- and cross-interactions among the various instability modes. We describe these in \S\ref{sec:optimal non-linear mechanisms for laminar--turbulent transition} by means of the non-linear \ac{STSM} optimization framework.

\section{Optimal non-linear mechanisms for laminar--turbulent transition\label{sec:optimal non-linear mechanisms for laminar--turbulent transition}}

In this section we analyse the physical mechanisms governing transition in the \ac{SWBLI}, from the onset of shear-layer development to breakdown. To ensure that the nonlinear interactions driving this process are properly resolved, we examine several \ac{STSM} truncations. The \ac{STSM} represents the solution in a finite Fourier space, with temporal and spanwise harmonics $n\in[-N,N]$ and $m\in[-M,M]$ (equivalently, $(2N+1)\times(2M+1)$ collocation points). This finite representation introduces truncation errors, which typically increase with forcing amplitude, $A=\|\mathbf{f}\|_F$, as nonlinear interactions transfer energy to higher-order harmonics. Adequate spectral resolution is therefore required to obtain converged transitional solutions \cite{rigas2021HBM,poulain2024,Savarino_Sipp_Rigas_2025}; the associated convergence study is reported in \S\ref{subsec:STSM system architectures}.

We adopt a \emph{fundamental} forcing configuration, following our earlier work on an incompressible laminar separation bubble~\cite{Savarino_Sipp_Rigas_2025}, in which the optimisation is restricted to the harmonics $(\pm1\omega,0\beta)$, $(0\omega,\pm1\beta)$ and $(\pm1\omega,\pm1\beta)$. These correspond, respectively, to 2D planar travelling waves, steady 3D disturbances (e.g. streamwise streaks/vortices), and 3D oblique waves. The forcing frequency $\omega$ and spanwise wavenumber $\beta$ are chosen from the most amplified linear instability identified by resolvent analysis at $(fX_{sh}/U_{\infty},\,\beta X_{sh})=(2,45)$ (first Mack-mode oblique waves; see \S\ref{appB}), in order to directly excite the oblique-wave branch in the quasi-linear regime. 

At the smallest forcing amplitude, $A=0.1\times10^{-5}$, the $(\pm1\omega,\pm1\beta)$ components are initialised using the corresponding resolvent forcing mode, while all other harmonics are set to zero. The nonlinear solutions are then obtained by continuation in $A$, using the converged forcing and response at the previous amplitude as the initial guess. The robustness of both the forcing configuration and the selected $(\omega,\beta)$ is assessed through two parametric studies in \S\ref{appC}.

\subsection{STSM system architectures}\label{subsec:STSM system architectures}
\begin{figure}
    \includegraphics[width=0.7\textwidth]{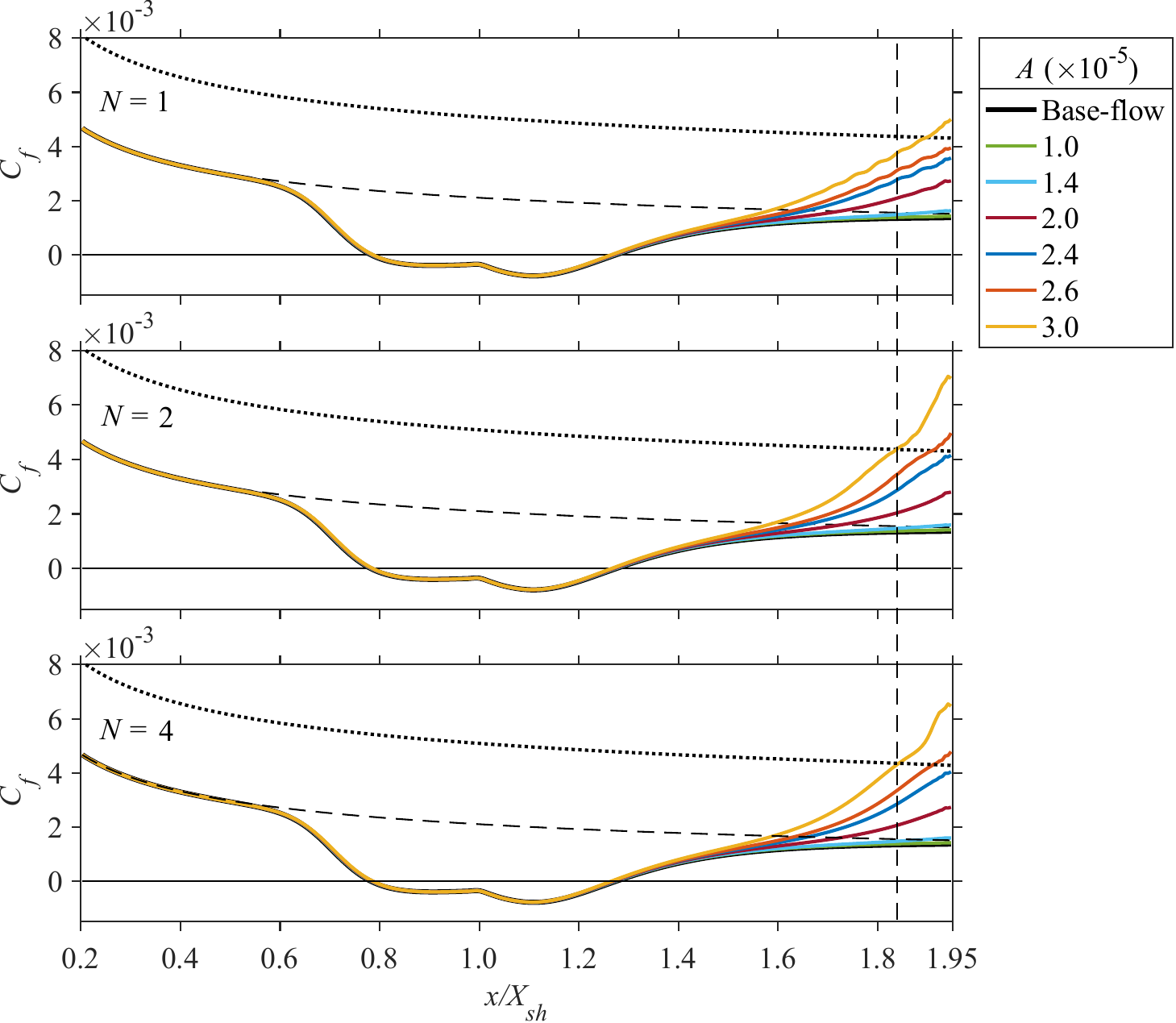}
    \caption{Mean skin friction coefficient calculated at different forcing amplitudes by systems (top) $N=1,M=4$, (middle) $N=2,M=4$ and (bottom) $N=4,M=4$, using the fundamental forcing configuration. Black-dashed line: laminar \ac{ZPG} boundary layer \cite{Howarth_1948}. Black-dotted line: turbulent \ac{ZPG} boundary layer \cite{Franko_Lele_2013}. Black-solid: laminar base-flow. The vertical black-dashed line indicates the converged location where the skin friction crosses the turbulent curve.}
    \label{fig:convergence Cf}
\end{figure}

The accuracy of the \ac{STSM} truncations is assessed by comparing the predicted mean skin-friction coefficient $C_f$, which directly enters the cost functional~\eqref{eq:cost function}. Guided by our previous studies of transitional supersonic boundary layers \cite{poulain2024}, we fix the spanwise truncation to $M=4$ (i.e. $m\in[-4,4]$) and vary the number of retained temporal harmonics $N$. Figure~\ref{fig:convergence Cf} compares the resulting $C_f$ distributions for $N=1$ (top), $N=2$ (middle) and $N=4$ (bottom), over forcing amplitudes ranging from $A=1.0\times10^{-5}$—for which mean-flow distortion is negligible—to $A=3.0\times10^{-5}$, where the flow exhibits transitional behaviour.

For $N=1$, the computed solutions exhibit aliasing for $A\geq 2\times10^{-5}$, indicating insufficient temporal resolution to represent the nonlinear transfers at these amplitudes with only one temporal harmonic. In the $C_f$ distributions, aliasing manifests as spurious waviness that becomes increasingly pronounced as the forcing amplitude is raised. This behaviour is consistent with under-resolved nonlinear interactions: energy that should populate higher temporal harmonics instead folds back onto the retained modes, producing non-physical contributions. Increasing the temporal truncation to $N=2$ and $N=4$ largely suppresses these artifacts and yields $C_f$ trajectories that converge towards the empirical turbulent level for a compressible \ac{ZPG} boundary layer \cite{Franko_Lele_2013}. 
These results show that the $N=2,M=4$ truncation offers a suitable accuracy--cost compromise for the present study. In particular, the negligible difference between the $N=2$ and $N=4$ predictions implies that harmonics at $(\pm3\omega,m\beta)$ and $(\pm4\omega,m\beta)$ are not required to reproduce the transition scenario considered here. Unless stated otherwise, we therefore use the $N=2,M=4$ system in the remainder of the paper.

To streamline the discussion of nonlinear solutions, we introduce the terminology used below to refer to disturbances of different order. The Fourier representation allows the harmonic coefficients $\hat{\mathbf{q}}_{n,m}$ to be ranked by their indices $(n,m)$. We refer to as \emph{first-generation} the set of harmonics for which $|n|\leq1$ and $|m|\leq1$ (excluding $(0,0)$), i.e. $(\pm1\omega,0\beta)$, $(0\omega,\pm1\beta)$ and $(\pm1\omega,\pm1\beta)$. \emph{Second-generation} disturbances are those for which $\max(|n|,|m|)=2$, and likewise \emph{third-} and \emph{fourth-generation} disturbances correspond to $\max(|n|,|m|)=3$ and $4$, respectively. This classification provides a compact language to describe the successive stages of mode generation and interaction throughout the transition process.

\begin{figure}
    \includegraphics[width=\textwidth]{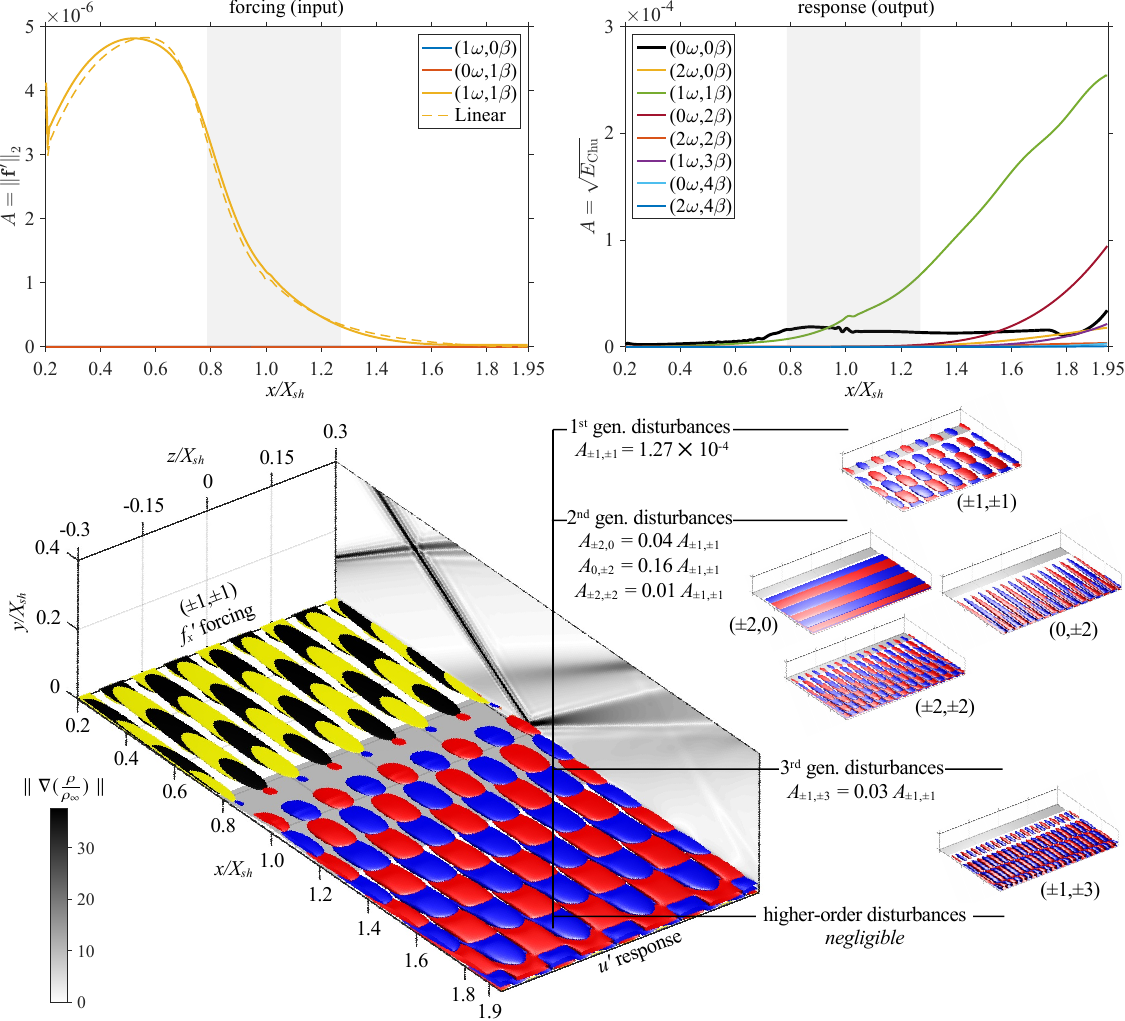}
    \caption{Optimal non-linear forcing/response solution at low amplitude $A=0.5\times10^{-5}$ computed from the $N=2,M=4$ system with the fundamental forcing configuration. Wall-normal integrated amplitudes of (top left) forcing harmonics based on the $L_2$-norm and (top right) response harmonics based on Chu's energy definition \cite{george2011chu}. The linear resolvent forcing mode is scaled to match the amplitude of the non-linear forcing for direct comparison. The mean separation length is plotted in gray. The $(0,0)$ harmonic is the mean-flow modification $\hat{\mathbf{q}}_{0,0}-\mathbf{q}_{0}$. (Bottom) isosurfaces of optimal oblique forcing ($f_{x}'$ component; yellow: positive, black: negative) and optimal response ($u'$ component; red: positive, blue: negative) obtained by summing all the harmonics except the mean-flow $\hat{\mathbf{q}}_{0,0}$. The small panels show the 3D structure ($u'$ isosurfaces) of the most energetic harmonics. The instantaneous separation bubble is plotted in gray. The $x$-$y$ plane at $z/X_{sh}=0.3$ shows the magnitude of the instantaneous first density gradient. The red-dashed line is the mean boundary layer displacement thickness.}
    \label{fig:forcing/response solution low amplitude}
\end{figure}

The optimal nonlinear forcing/response pair at the low amplitude $A=0.5\times10^{-5}$ is shown in Fig.~\ref{fig:forcing/response solution low amplitude}. At this amplitude the flow remains far from transition: the mean skin-friction coefficient is essentially indistinguishable from the laminar base-flow. Nevertheless, the response already displays a weakly nonlinear (\ac{WNL}) character, providing a clean setting to isolate the first nonlinear mechanism that follows the linear amplification stage.

The computed optimal forcing is entirely supported by the oblique-wave harmonic $(1\omega,1\beta)$ (Fig.~\ref{fig:forcing/response solution low amplitude}, top left), with its maximum amplitude located upstream of the separation point, in agreement with the linear resolvent prediction (see \S\ref{appB}). The resemblance between the rescaled linear optimal forcing and the nonlinear optimal forcing is striking. Although the fundamental forcing configuration permits contributions from $(1\omega,0\beta)$ and $(0\omega,1\beta)$, none are selected by the optimisation. This confirms that, in the present configuration, the oblique first Mack mode constitutes the dominant primary mechanism of the separated shear layer, consistent with the view of the shear layer as a selective disturbance amplifier that preferentially promotes oblique waves \cite{bugeat_robinet_chassaing_sagaut_2022,Mauriello_Sharma_Larchevêque_Sandham_2025,Saidi_Wang_Fournier_Tenaud_Robinet_2025}.

Consistent with the forcing, the $(1\omega,1\beta)$ response is seeded upstream of separation and subsequently amplifies over the separated shear layer (Fig.~\ref{fig:forcing/response solution low amplitude}, top right). The near-exclusive presence of this mode in the upstream portion of the interaction region indicates that the early shear-layer evolution is governed by essentially linear dynamics. The reconstructed disturbance field $u'$ is shown in Fig.~\ref{fig:forcing/response solution low amplitude} (bottom): the shear layer sheds large-scale wave packets with a characteristic checkerboard pattern, which we attribute to the oblique first Mack-mode instability. Although the reconstruction includes all non-zero harmonics, the overall response remains dominated by the fundamental oblique component, and the wave packets exhibit only limited streamwise elongation downstream of reattachment.

The side panels in Fig.~\ref{fig:forcing/response solution low amplitude} (bottom) further quantify the nonlinear content: the most energetic second-generation component reaches only $16\%$ of the energy of the first-generation oblique waves, while all remaining harmonics contribute less than $5\%$. This \ac{WNL} regime therefore provides a minimal model of primary-instability saturation and the first triadic interaction, which seeds the second-generation steady mode $(0\omega,2\beta)$ in the vicinity of mean reattachment. Closely related scenarios have been reported in \cite{Hildebrand2018,dwivedi_sidharth_jovanovic_2022} and are revisited here in \S\ref{subsec:WNL stage: quadratic interaction seeding Görtler vortices and streaks} within the nonlinear input--output framework. A key ingredient of this analysis is the identification of the nonlinear coupling terms of the compressible \ac{N-S} equations from which the intrinsic forcing mechanisms are extracted (see \S\ref{subsec:calculation of non-linear couplings in the compressible Navier-Stokes}).

\subsection{Calculation of nonlinear couplings in the compressible Navier--Stokes}
\label{subsec:calculation of non-linear couplings in the compressible Navier-Stokes}

The compressible \ac{N-S} equations contain nonlinear couplings of different polynomial order in the conservative variables. Here we focus on the convective momentum flux,
$\mathcal{F}_{\mathrm{conv.}}=\rho\,\mathbf{u}\mathbf{u}$, in the momentum equations~\eqref{subeq:momentum}, since it provides a direct mechanism by which existing disturbances generate new harmonics through intrinsic nonlinear interactions. While analogous contributions also arise from the convective enthalpy flux and viscous dissipation in the energy equation~\eqref{subeq:energy}, we restrict attention to momentum transport, which suffices to elucidate the dominant coupling pathways in the transition scenarios considered.

Using Einstein notation, the convective momentum flux reads $(\mathcal{F}_{\mathrm{conv.}})_{ij}=\rho\,u_i u_j$ with $i,j\in\{1,2,3\}$. Decomposing the density and velocity into mean and fluctuating components,
$\rho=\overline{\rho}+\rho'$ and $u_i=\overline{u}_i+u_i'$, and expanding the product yields the fluctuation-induced contributions to the momentum flux. Grouping terms by order in the fluctuations gives
\begin{align}
(\mathcal{F}^{(2)}_{\mathrm{conv.}})_{ij} &=
\overline{\rho}\,u_i' u_j'
+\rho' u_i' \overline{u}_j
+\rho'\,\overline{u}_i u_j',
\label{eq:quadratic coupling}\\[4pt]
(\mathcal{F}^{(3)}_{\mathrm{conv.}})_{ij} &=
\rho'\,u_i' u_j',
\label{eq:cubic coupling}
\end{align}
where the overbar denotes the $(t,z)$-mean and the prime denotes the deviation from this mean.

These nonlinear fluxes can be interpreted as an \emph{intrinsic forcing} of the linearised momentum equations. Specifically, taking the divergence and moving the resulting term to the right-hand side defines the three-component forcing vector
\begin{equation}
\mathbf{f}_{\mathrm{conv.}}
= -\nabla\cdot\!\left(\mathcal{F}^{(2)}_{\mathrm{conv.}}+\mathcal{F}^{(3)}_{\mathrm{conv.}}\right),
\label{eq:intrinsic forcing}
\end{equation}
which acts only on the momentum equations by construction.

To obtain the contribution of this intrinsic forcing at a specific harmonic $(n,m)$, we compute its Fourier coefficient $\hat{\mathbf{f}}_{n,m}$ by applying the \ac{DFT} of~\eqref{eq:intrinsic forcing} over $(t,z)$ (equivalently, over the corresponding collocation grid in the \ac{STSM}). This makes the triadic structure of the interactions explicit: for instance, the $(0,2)$ forcing discussed in \S\ref{subsec:WNL stage: quadratic interaction seeding Görtler vortices and streaks} arises from the quadratic interaction $(1,1)+(-1,1)$, whereas the $(1,3)$ forcing in \S\ref{subsec:highly non-linear stage: secondary streak instability} may result from either a quadratic pathway, $(1,1)+(0,2)$, or a cubic pathway, $(1,1)+(1,1)+(-1,1)$. In the next two subsections we analyse these interaction pathways and relate them to the physical mechanisms that drive the successive stages of transition.

\subsection{Weakly nonlinear stage: quadratic interaction seeding G\"ortler vortices and streaks}
\label{subsec:WNL stage: quadratic interaction seeding Görtler vortices and streaks}

At $A=0.5\times10^{-5}$ (Fig.~\ref{fig:forcing/response solution low amplitude}), the optimal response remains dominated by the oblique first Mack-mode waves, while a secondary three-dimensional steady component becomes detectable downstream of mean reattachment. In this weakly nonlinear regime, the leading mechanism responsible for generating higher-order content is the quadratic convective coupling~\eqref{eq:quadratic coupling}. In particular, the interaction of two oblique Mack-wave harmonics, $(1,1)$ and $(-1,1)$, produces a steady spanwise harmonic at $(0,2)$, thereby seeding the $(0,2)$ response observed in the \ac{WNL} solution. By contrast, cubic pathways involving only first-generation waves (e.g. $(1,1)+(1,1)+(-1,1)\rightarrow(1,3)$) are found to be substantially weaker at this amplitude.

\begin{figure}
    \includegraphics[width=\textwidth]{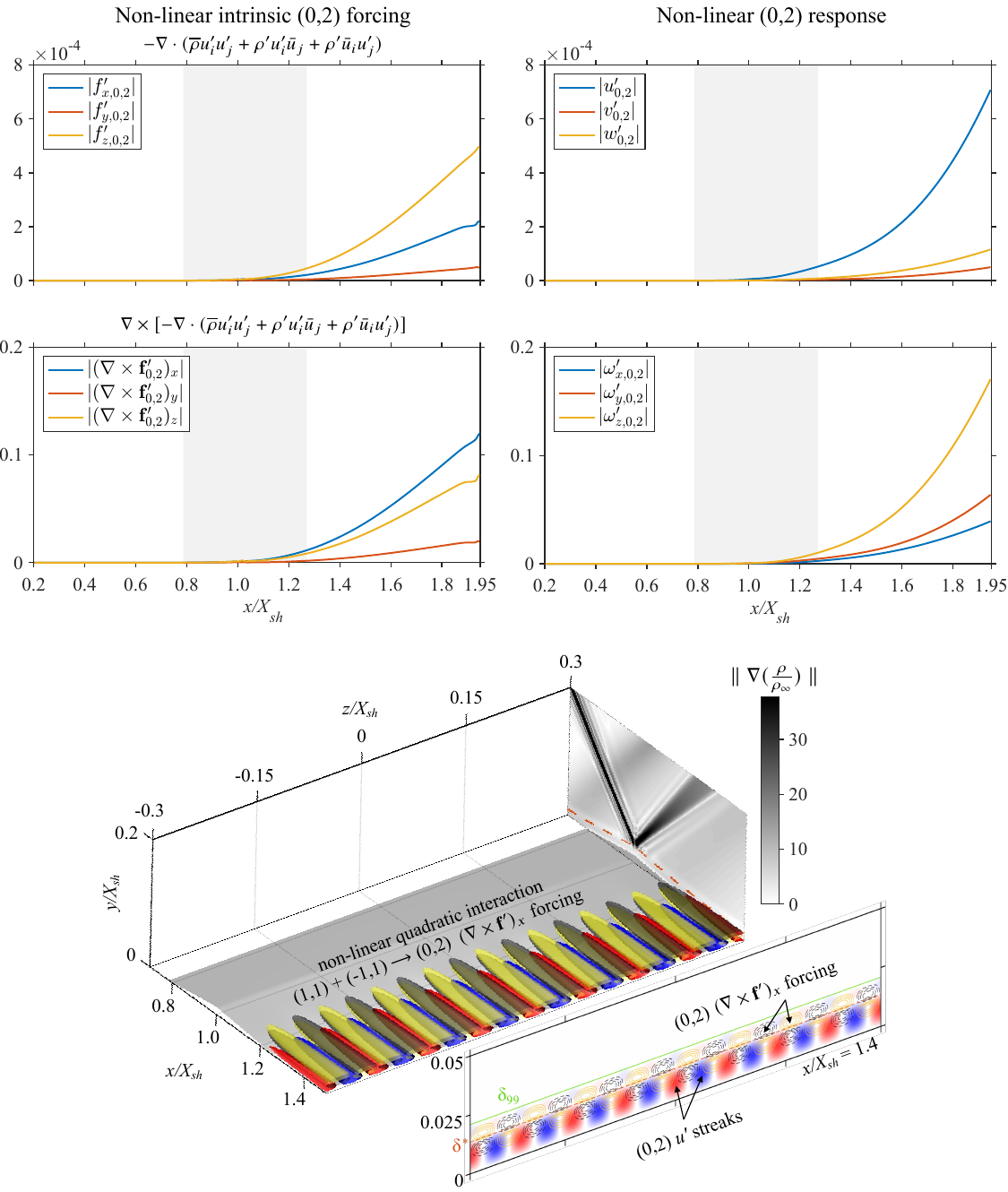}
    \caption{Non-linear quadratic (0,2) mechanism at low amplitude $A=0.5\times10^{-5}$. Top left: component-wise amplitudes of non-linear forcing. Top right: component-wise velocity disturbance amplitudes of non-linear (0,2) response. Middle left: component-wise amplitudes of the curl of non-linear forcing. Middle right: component-wise vorticity disturbance amplitudes of non-linear (0,2) response. Bottom: isosurfaces of streamwise vortical (0,2) forcing (yellow: positive, black: negative) and streamwise velocity (0,2) response (red: positive, blue: negative) superimposed on the instantaneous separation bubble. The $x$-$y$ plane at $z/X_{sh}=0.3$ shows the magnitude of the instantaneous first density gradient. A $z$-$y$ planar view of the (0,2) disturbances is extracted at $x/X_{sh}=1.4$.}
    \label{fig:(0,2) forcing-response mechanism}
\end{figure}

We study the quadratic mechanism in figure \ref{fig:(0,2) forcing-response mechanism}. The top left panels show the component-wise amplitudes of the non-linear (0,2) forcing and its curl (to be interpreted as vorticity) produced by the quadratic non-linearity. While the panels display the cumulative contribution of the density-velocity couplings and the Reynolds stresses, the former are of negligible significance, meaning the mechanism under investigation is essentially driven by velocity fluctuations. The leading component of the forcing curl is in the streamwise direction, fed by large wall-normal gradients of the spanwise forcing disturbance. The 3D structure of these streamwise vortical excitations is reconstructed in the bottom panel of figure \ref{fig:(0,2) forcing-response mechanism}, showing an arrangement of streamwise-oriented vortices of alternating sign located at the mean boundary layer displacement thickness height. As thoroughly discussed in \S\ref{appD} and in existing \ac{SWBLI} literature \cite{loginov_adams_zheltovodov_2006,kontis_erdem_johnstone_murray_steelant_2013,kuehl_paredes_2016,roghelia_olivier_egorov_chuvakhov_2017,Pasquariello_Hickel_Adams_2017,Giepman_Schrijer_vanOudheusden_2018,shinde2019,currao_2020,currao_2021,zhao_ma_chen_zhang_hao_wen_2024,jiao_ma_xue_wang_chen_cheng_2024,dixit_kumar_vadlamani_tsuboi_2025,sun_yu_li_zhang_2025}, this disturbance is reminiscent of Görtler-type vortices, a convective instability that arises due to the imbalance between centrifugal and pressure forces in the direction normal to curved streamlines \cite{Gortler1941,floryan_saric_1982,hall_malik_1989,floryan1991,Saric1994,li_malik_1995,Ren2018}. Therefore, the conditions of local streamline curvature and non-linear excitations produced by the quadratic interaction of Mack waves, make the mean reattachment region a susceptive site for the development of Görtler vortices. The present scenario agrees with other studies that have reported the emergence of low-frequency/quasi-steady 3D disturbances seeded by medium-frequency (shear layer) fluctuations due to non-linear effects \cite{dwivedi_sidharth_jovanovic_2022,Saidi_Wang_Fournier_Tenaud_Robinet_2025,Mauriello_Sharma_Larchevêque_Sandham_2025}.

\begin{figure}
    \includegraphics[width=0.8\textwidth]{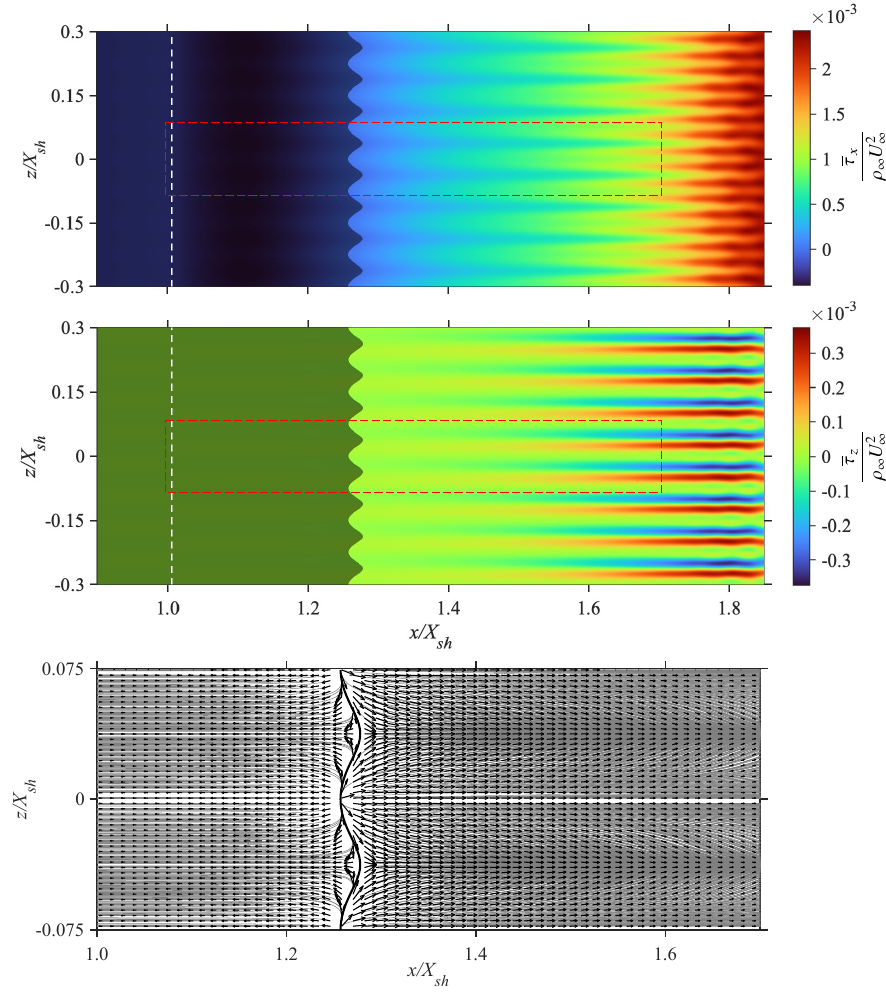}
    \caption{Trace of Görtler-like vortices in the time-averaged flow field at amplitude $A=3.0\times10^{-5}$. Top: time-averaged streamwise wall shear stress. Middle: time-averaged spanwise wall shear stress. The time-averaged separation bubble isosurface is superimposed. The white-dashed line marks the mean bubble apex. Bottom: skin friction lines at the wall and vectors from the wall shear vector field $(\overline{\tau}_x,\overline{\tau}_z)$ plotted within the part of the domain highlighted with the red-dashed box. The black-solid line indicates the time-averaged reattachment.}
    \label{fig:wall shear}
\end{figure}

Velocity streaks are generated by Görtler-like vortices since, at the spanwise wavenumber of the present centrifugal instability mode $\beta X_{sh}=90$, the \ac{SWBLI} flow is mildly receptive to streak-type instabilities (see figure \ref{fig:resolvent gain} in \ref{appB}). This is shown in figure \ref{fig:(0,2) forcing-response mechanism} (top right), where the disturbance velocity field is dominated by $u'$ (the streak part) and the vorticity field contains a non-zero trace of $\omega_{x}'$ (the centrifugal instability part) that is negligible in \ac{ZPG} boundary layer (refer to the works of \cite{rigas2021HBM,poulain2024,Savarino_Sipp_Rigas_2025}). High- and low-speed streaks are also represented three-dimensionally in the bottom panel. The characteristic pattern stems from the upwash/downwash transport of streamwise momentum across the boundary layer imparted by the counter-rotating Görtler-like vortices \cite{shinde2019,zhao_ma_chen_zhang_hao_wen_2024,chen2024,sun_yu_li_zhang_2025}. This mechanism is reminiscent of lift-up observed in both incompressible and compressible boundary layers at the early stages of transition \cite{schmid1992new,waleffe1997, rigas2021HBM,bugeat_robinet_chassaing_sagaut_2022,poulain2024}. In regions of downwash $u'>0$ streaks appear, conversely, regions of upwash give $u'<0$ streaks. This process is also responsible for the corrugated reattachment front of the mean separation bubble, as illustrated in figure \ref{fig:wall shear}. The $u=0$ isosurface representing the 3D structure of the bubble shows that spanwise corrugations appear on the rear side of the separation zone, downstream of the mean bubble apex, matching the location where the streamwise vortices are originally seeded by the the quadratic non-linearity. In contrast, the front side is essentially 2D with no visible spanwise modulations. The wavy pattern of the reattachment line follow the spatial arrangement of streamwise vorticity, in that two adjacent vortices of opposite rotation entraining high streamwise momentum fluid from the outer shear layer into the inner layers (downwash region) enhance the inertial forces opposing to the \ac{APG}, promoting shear layer reattachment (for example at $z/X_{sh}=0$ in figure \ref{fig:wall shear}, bottom). On the contrary, the uplift of low streamwise momentum fluid from the inner layers into the shear layer by another vortex pair reduces the inertia of the shear layer, meaning a longer distance in the streamwise direction is required to fight the \ac{APG} and reattach the flow (see location $z/X_{sh}$ midway between 0 and 0.075 in figure \ref{fig:wall shear}, bottom). 

The presence of Görtler-like vortices can also be detected by inspecting the mean wall shear $(\overline{\tau}_{x},\overline{\tau}_{z})$. Striations in $\overline{\tau}_{x}$ begin to develop from the reattachment region with the same spanwise wavelength ($\lambda_{z}=5$-$6~\mathrm{mm}$) of the streamwise vortices (figure \ref{fig:wall shear}, top). High/low shear is present in regions of downwash/upwash. Similar observations for the wall shear by \cite{loginov_adams_zheltovodov_2006,Pasquariello_Hickel_Adams_2017,sun_yu_li_zhang_2025}, the Stanton number (heat flux) by \cite{kontis_erdem_johnstone_murray_steelant_2013,roghelia_olivier_egorov_chuvakhov_2017,currao_2020,cao_hao_klioutchnikov_olivier_wen_2021,currao_2021,jiao_ma_xue_wang_chen_cheng_2024,dixit_kumar_vadlamani_tsuboi_2025} and the wall temperature distribution by \cite{baskaya_dungan_hickel_brehm_2024} attest the origin of these characteristic spanwise striations from Görtler vortices, which become particularly critical in \ac{SWBLI}s wherein peak heating spots may endanger the structural health of materials \cite{smiths_dussauge_2006,babinsky_harvey_2011,gaitonde2015progress}. High/low streamwise wall shear streaks also appear in flat plate \ac{ZPG} boundary layers as a result of lift-up (see figure \ref{fig:wall shear ZPG}, top, in \S\ref{appE}), although with different magnitude compared to \ac{SWBLI}. These streaks are therefore not exclusive to Görtler instability since $\overline{\tau}_{x}$ is related to $\partial \overline{u}/\partial y$, which is also a consequence of lift-up. It is therefore worth considering the spanwise component of the wall shear $\overline{\tau}_{z}$ in figure \ref{fig:wall shear} (middle). Streak pairs containing two streaks of opposite sign grow from the reattachment region in the downstream direction due to the action of counter-rotating vortices creating $\overline{w}$ gradients. These striations are mild in \ac{ZPG} boundary layer until streak breakdown occurring at $x/X_{sh}\approx1.8$ (figure \ref{fig:wall shear ZPG}, middle). In the \ac{SWBLI}, spanwise shear is stronger in the post-reattachment region, due to Görtler vortex development.

Equipped with the $(\overline{\tau}_{x},\overline{\tau}_{z})$ field, we can study the topology of the flow near the wall by plotting the skin friction lines, which are the solution to the equation set $\mathrm{d}x/\mathrm{d}s=\overline{\tau}_{x}(x,z),\:\mathrm{d}z/\mathrm{d}s=\overline{\tau}_{z}(x,z)$ \cite{delery2001,knight2003advances}. These are shown juxtaposed to vectors of the $(\overline{\tau}_{x},\overline{\tau}_{z})$ field in figure \ref{fig:wall shear} (bottom). Notably, skin friction lines gets suddenly bent towards the center of each streak pair in the $\overline{\tau}_{z}$ contours just downstream of the mean reattachment point, a feature that is not observed in \ac{ZPG} boundary layer in the same way (figure \ref{fig:wall shear ZPG}, bottom). In fact, the stronger $\overline{\tau}_{z}$ in the reattachment zone of the \ac{SWBLI} yields skin friction lines with more pronounced inclination, ultimately ascribed to Görtler vortices.

Depending on the spanwise wavelength, the Görtler number commonly used as a measure of strength of the vortices and the Mach number, Görtler vortices can themselves breakdown and lead to turbulence via different instability modes \cite{Saric1994,li_malik_1995,schrader_brandt_zaki_2011,li_choudhari_chang_greene_wu_2012,Xu_Zhang_Wu_2017,roghelia_olivier_egorov_chuvakhov_2017,souza_2017,li_choudhari_paredes_2022,xu_ricco_duan_2024}. Alternatively, they can act as transitional instabilities that can seed other late-stage mechanisms for breakdown (like streak breakdown or \ac{T-S}/Mack wave breakdown) \cite{Saric1994,li_malik_1995,chen_chen_yuan_tu_zhang_2019,kuehl_paredes_2016}. This new stage in the transition process is explored in \S\ref{subsec:highly non-linear stage: secondary streak instability}.

\subsection{Highly nonlinear stage: secondary streak instability}
\label{subsec:highly non-linear stage: secondary streak instability}

The late transitional regime preceding turbulent breakdown is considered. In particular, the solution at $A=3.0\times10^{-5}$ is analysed, for which the mean skin-friction coefficient exhibits a clear rise to transitional levels (see Fig.~\ref{fig:convergence Cf}, middle), indicating that transition has occurred.

As suggested by the Görtler analysis in \S\ref{appD}, the maximum Görtler number $G_{T}$ and curvature parameter $\delta/\mathcal{R}$ computed at the reattachment point are sufficiently high to support unstable Görtler vortices according to the criteria of \cite{Gortler1941,floryan1991,smiths_dussauge_2006}, but are comparatively smaller than typical values found in \ac{SWBLI} literature where transition is driven by secondary instability and breakdown of Görtler vortices \cite{loginov_adams_zheltovodov_2006,roghelia_olivier_egorov_chuvakhov_2017,currao_2020,currao_2021,dixit_kumar_vadlamani_tsuboi_2025}. To gain physical insight, we study the role of the (1,3) response mode in the development of 3D flow structures at high amplitude and compare with the \ac{ZPG} boundary layer benchmark. Furthermore, new triadic interactions producing forcing at (1,3) are calculated, as previously done in \S\ref{subsec:WNL stage: quadratic interaction seeding Görtler vortices and streaks} for the (0,2) mode.

\begin{figure}
    \includegraphics[width=\textwidth]{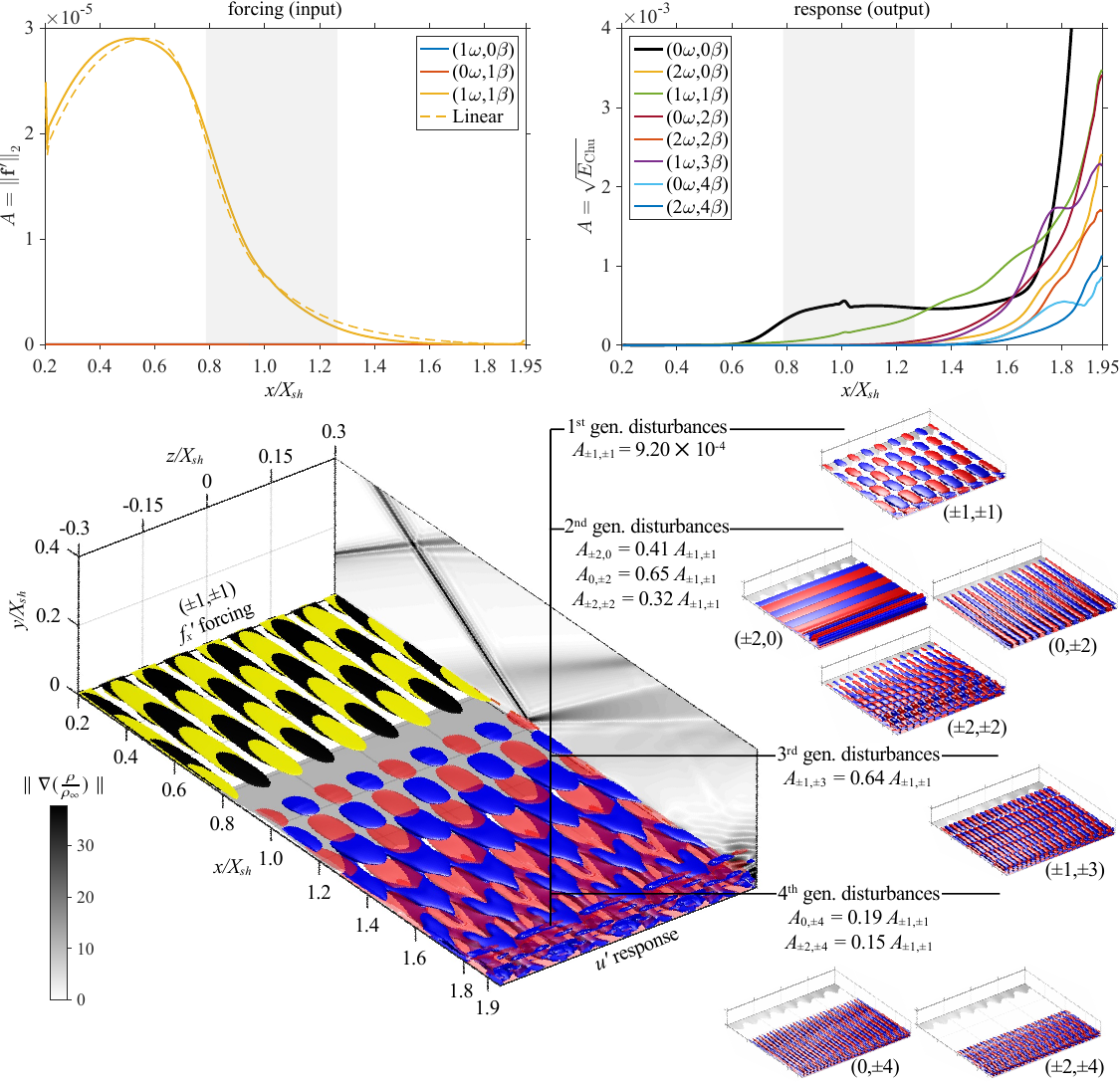}
    \caption{Optimal forcing/response solution at high amplitude $A=3.0\times10^{-5}$ computed from the $N=2,M=4$ system with the fundamental forcing configuration. The same quantities as in figure \ref{fig:forcing/response solution low amplitude} are plotted. The positive $u'$ (red) isosurfaces of the perturbation flow reconstruction are shaded to emphasize the $\Lambda$-shaped structures appearing in the negative (blue) isosurfaces.}
    \label{fig:forcing/response solution high amplitude}
\end{figure}

The high amplitude/transitional \ac{SWBLI} solution is shown in figure \ref{fig:forcing/response solution high amplitude}. By comparison with the low amplitude case in figure \ref{fig:forcing/response solution low amplitude}, the reconstructed $u'$ flow field develops strongly 3D, $\Lambda$-type structures typical of late transitional stages. The response solution (figure \ref{fig:forcing/response solution high amplitude}, top right) is now populated by a number of disturbances that are non-linearly excited in the reattached boundary layer region. Most importantly, the amplitude of the 3rd generation harmonic (1,3) becomes nearly 65\% of the first generation (1,1) mode, similarly to 2nd generation (0,2). On the other hand, the optimal (1,1) forcing remains spatially and structurally unchanged with respect to the low amplitude solution, suggesting that transition can be achieved with the same type of oblique forcing since, once the primary Mack instability is excited, the intrinsic non-linear mechanisms of the flow trigger laminar--turbulent transition.

\begin{figure}
    \includegraphics[width=0.85\textwidth]{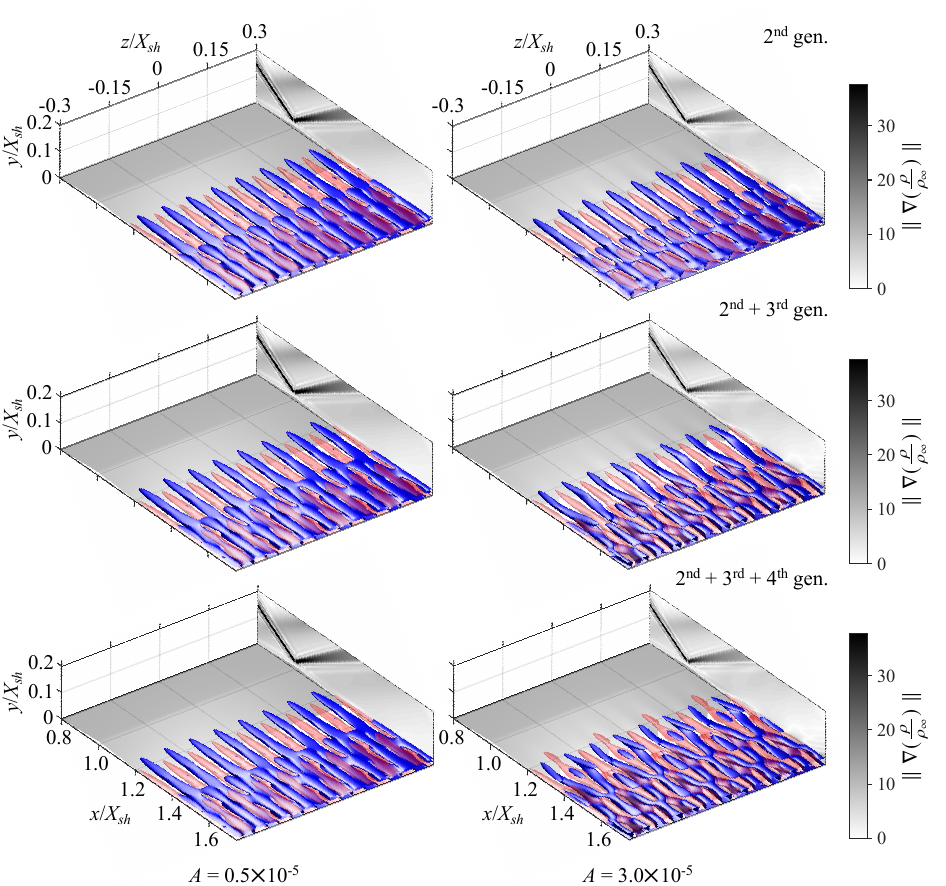}
    \caption{Streak instability via (1,3) harmonic mode. (Left) low amplitude $A=0.5\times10^{-5}$. (Right) high amplitude $A=3.0\times10^{-5}$. Isosurfaces of $u'$ (red: positive, blue: negative) reconstructed with (top) 2nd, (middle) 2nd and 3rd, (bottom) 2nd, 3rd and 4th generation disturbances. The instantaneous separation bubble isosurface is superimposed. The $x$-$y$ plane at $z/X_{sh}=0.3$ shows the magnitude of the instantaneous first density gradient.}
    \label{fig:streak instability}
\end{figure}

The influence of the $(1,3)$ disturbance on the reconstructed flow topology is examined in Fig.~\ref{fig:streak instability}, motivated by its three-dimensional character and by the presence of $\Lambda$-type structures in the post-reattachment boundary layer. The streamwise perturbation velocity $u'$ is shown in the post-reattachment region using (top) second-generation harmonics only, (middle) second- and third-generation harmonics, and (bottom) second- to fourth-generation harmonics, for both the (left) low-amplitude and (right) high-amplitude cases. To isolate the ``streaky'' component of the flow, the first-generation $(1,1)$ mode is intentionally omitted, thereby removing the dominant oblique-wave pattern.
In the \ac{WNL} case, the reconstructions are essentially indistinguishable across truncations: the low-speed streaks remain largely straight and no signature of secondary instability is apparent. In contrast, at high amplitude, the inclusion of the third-generation $(1,3)$ harmonic on top of the second-generation $(0,2)$ component introduces a pronounced sinuous deformation of the low-speed streaks (middle right panel). This pattern is consistent with the subharmonic sinuous secondary instability reported for streaks in \ac{ZPG} boundary layers \cite{andersson_brandt_bottaro_henningson_2001} (see Fig.~\ref{fig:streak instability ZPG} for a qualitative comparison). Upon further including the fourth-generation harmonics, additional smaller-scale three-dimensional structures become visible, while the large-scale sinuous motion of the streaks remains clearly identifiable.

\begin{table}
    \caption{$L_2$-norm based amplitude of intrinsic (1,3) forcing contributions from quadratic and cubic interactions at forcing amplitude $A=3.0\times10^{-5}$.}
    \centering
    \begin{tabular}{ccccccc}
    \hline
    $\mathbf{Order}$ & $\boldsymbol{(n_{1},m_{1})}$ & $\boldsymbol{(n_{2},m_{2})}$ & $\boldsymbol{(n_{3},m_{3})}$ & $\mathbf{Component}$ $\boldsymbol{(-\nabla\cdot)}$ & $\boldsymbol{A}$ $\mathbf{(L_2}$-$\mathbf{norm})$ & $\boldsymbol{A/A_{\mathrm{tot}}}$ $\mathbf{(\%)}$ \\\hline
    Quadratic & (1,1) & (0,2) & $-$ & $\overline{\rho}u_{i}'u_{j}'$ & 0.00761 & 47.8\\
     & & & & $\rho'u_{i}'\overline{u}_{j}+\rho'\overline{u}_{i}u_{j}'$ & 0.00078 & 4.9\\
    Quadratic & (-1,1) & (2,2) & $-$ & $\overline{\rho}u_{i}'u_{j}'$ & 0.00654 & 41.1\\
     & & & & $\rho'u_{i}'\overline{u}_{j}+\rho'\overline{u}_{i}u_{j}'$ & 0.00052 & 3.2\\
    Cubic & (1,1) & (1,1) & (-1,1) & $\rho'u_{i}'u_{j}'$ & 0.00023 & 1.4\\
    Cubic & (1,1) & (2,0) & (-2,2) & $\rho'u_{i}'u_{j}'$ & 0.00012 & 0.8\\
    Cubic & (1,1) & (-2,0) & (2,2) & $\rho'u_{i}'u_{j}'$ & 0.00012 & 0.8\\
    \hline
    \end{tabular}
    \label{tab:(1,3) interactions}
\end{table}

The nonlinear pathways that force the $(1,3)$ response are summarised in Table~\ref{tab:(1,3) interactions}. Five distinct triadic combinations can, in principle, generate intrinsic forcing at the $(1,3)$ harmonic. Their relative importance is quantified using the fraction of the total $L_2$-norm amplitude of the $(1,3)$ forcing vector, denoted $A_{\mathrm{tot}}$. The dominant contribution arises from the quadratic interaction between first- and second-generation harmonics, $(1,1)+(0,2)$, followed by $(-1,1)+(2,2)$. As for the $(0,2)$ mechanism discussed in \S\ref{subsec:WNL stage: quadratic interaction seeding Görtler vortices and streaks}, density-related terms are found to be negligible, indicating that the forcing is primarily produced by velocity-fluctuation interactions through the Reynolds-stress contribution $\overline{\rho}\,u_i' u_j'$. Cubic pathways involving only first-generation disturbances or mixed first-/second-generation interactions are approximately an order of magnitude weaker than the leading quadratic terms and contribute negligibly to $A_{\mathrm{tot}}$.

Overall, the prevalence of the $(1,1)+(0,2)\rightarrow(1,3)$ pathway suggests a secondary streak-instability mechanism analogous to that reported in \ac{ZPG} boundary layers, in both incompressible \cite{schmid1992new,andersson_brandt_bottaro_henningson_2001,rigas2021HBM} and compressible \cite{poulain2024} settings. This observation supports the view that, once reattached, the shear layer downstream of the shock-induced separation recovers transition dynamics that closely resemble those of canonical boundary layers.

\begin{figure}
    \includegraphics[width=\textwidth]{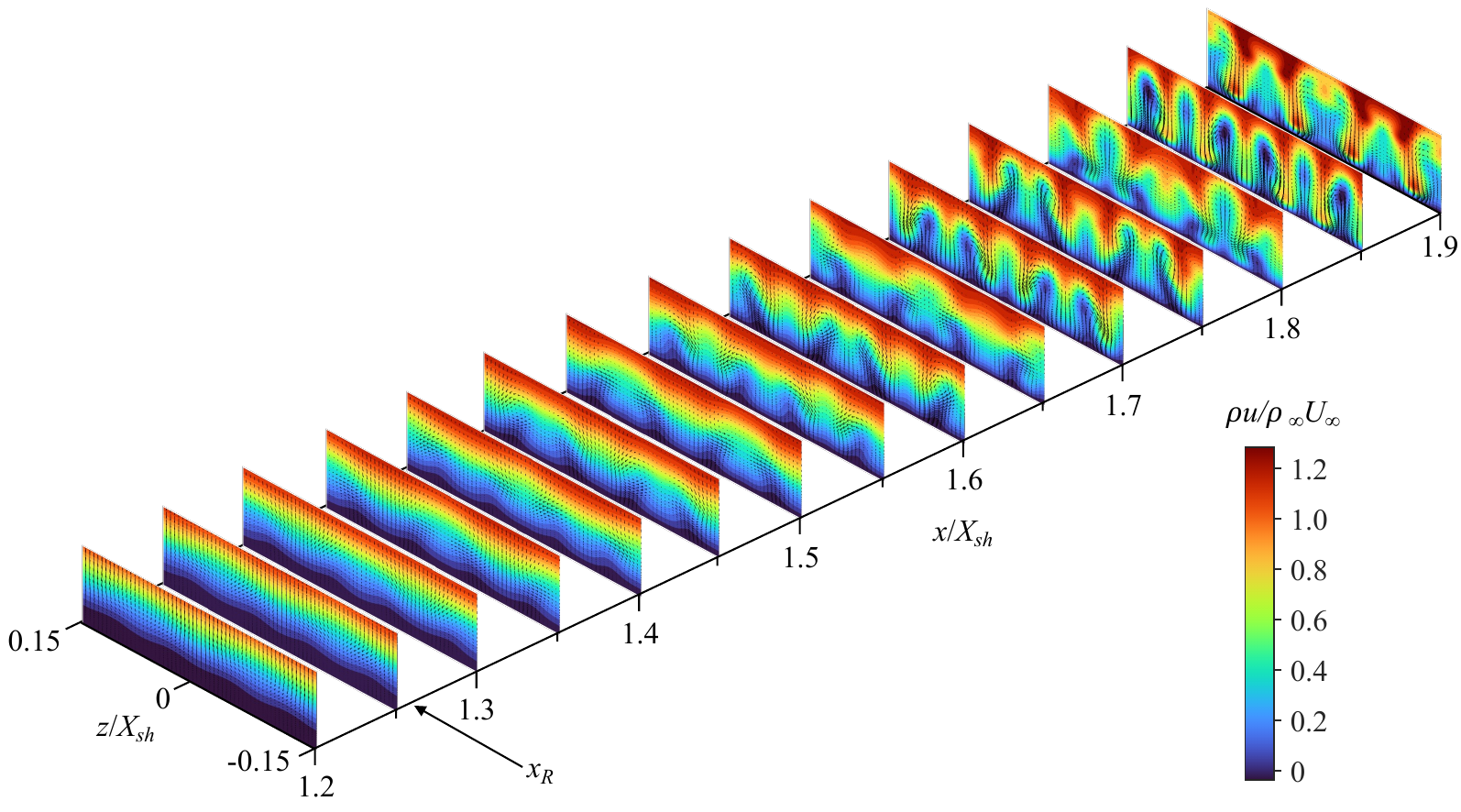}
    \caption{Contours of streamwise momentum $\rho u$ and $(v,w)$ velocity vectors of the instantaneous flow field at several streamwise stations showing the progressive 3D distortion of the boundary layer in the transition process. Solution obtained at amplitude $A=3.0\times10^{-5}$.}
    \label{fig:velocity contours}
\end{figure}

At $x/X_{sh}=1.6$ in Fig.~\ref{fig:velocity contours}, enhanced cross-stream mixing of streamwise momentum is observed concurrently with the emergence of $\Lambda$-shaped structures \cite{schmid1992new,rigas2021HBM,dwivedi_sidharth_jovanovic_2022,poulain2024,Savarino_Sipp_Rigas_2025}. Further cross-stream visualisations downstream of this station reveal pronounced upwash of low-speed fluid, driven by streamwise vortical motions as indicated by the $(v,w)$ velocity vectors. Beyond $x/X_{sh}=1.7$, the boundary layer becomes strongly three-dimensional and smaller-scale motions appear; these are also evident in the $Q$-criterion isosurfaces in Fig.~\ref{fig:overview}. Collectively, these features are interpreted as precursors to turbulent breakdown.

\section{\label{sec:conclusions}Concluding remarks}

The nonlinear evolution of a convectively unstable, shock-induced transitional shear layer subjected to environmental disturbances was characterised using the \ac{HBNS} input--output optimisation framework applied to the compressible \ac{N-S} equations \cite{rigas2021HBM,poulain2022,poulain2024}.

A separated \ac{SWBLI} configuration from the experimental test case of \cite{degrez1987interaction} was simulated numerically. The laminar base-flow obtained for an incident shock angle $\theta=30.8^{\circ}$ displayed a globally stable separation bubble (sustaining no self-excited resonances) and a convectively unstable/noise-amplifier separated shear layer, making this a suitable case for the study of external disturbance/noise-driven transition to turbulence.

While the linearized dynamics of disturbances via the resolvent-input/output and global eigen-problem frameworks revealed a number of different linear instability mechanisms, such as oblique waves (via the first Mack instability mode), streaks (due to lift-up) and a very damped (for $\theta=30.8^{\circ}$) modal resonance ascribed to the low-frequency bubble breathing mechanism, the analysis of the transition pathway to turbulence, characterized by the synergistic action and interaction of these disturbances, was made possible by the compressible extension of the \ac{AHBM}: the \ac{STSM} \cite{poulain2024}. The second main objective of the presented work was to identify ``worst-case'' conditions under which the shock-induced shear layer would transition to turbulence with minimal energy input from external disturbances, thus elucidating efficient mechanisms of turbulence production exploited by the inherent non-linear dynamics of the flow.

\begin{figure}
    \includegraphics[width=\textwidth]{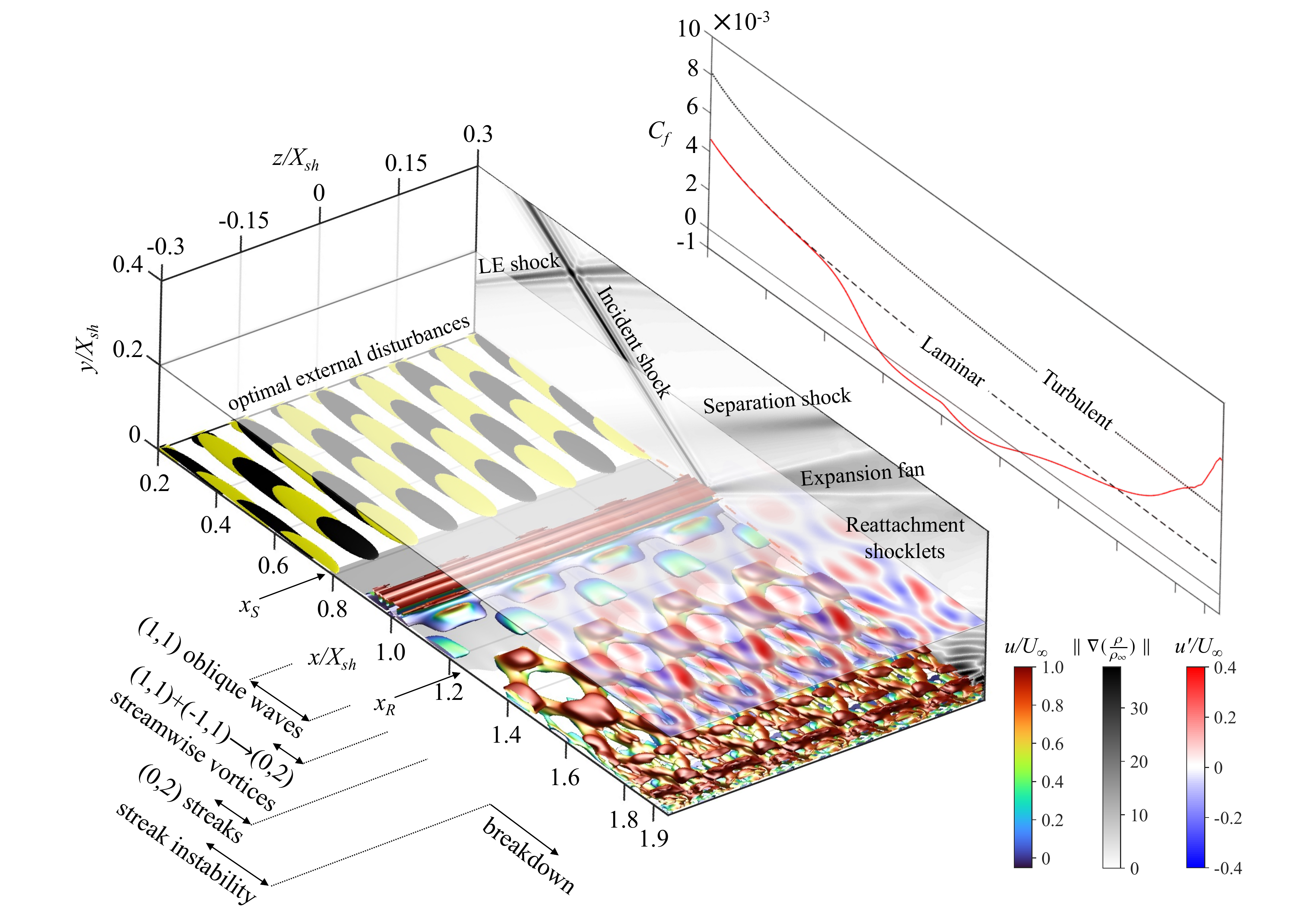}
    \caption{Overview of the optimal route to turbulence of the \ac{SWBLI} forced by external disturbances. (Left) instantaneous 3D flow (external disturbances, flow field visualization via Q-criterion, separation bubble, planar view of $u'$ fluctuations and magnitude of the first density gradient). (Right) mean skin friction coefficient. Solution obtained at amplitude $A=3.0\times10^{-5}$.}
    \label{fig:overview}
\end{figure}

The identified optimal route to turbulence of the \ac{SWBLI} is summarized visually in figure \ref{fig:overview} and is broken down into few stages as below:
\begin{enumerate}
    \item Oblique wave-like disturbances optimally forcing the shear layer upstream of separation lead to the amplification of the first Mack mode through the receptivity process at the selective frequency range $fX_{sh}/U_{\infty}\sim10^{0}$, a process well described by linear resolvent at small disturbance amplitude.
    \item At finite amplitude, the non-linear (quadratic) self-interaction of (1,1) Mack waves seeds (0,2) streamwise Görtler-like vortices in the reattachment region where the streamline curvature is sufficient to support unstable Görtler modes.
    \item Velocity (0,2) streaks are generated by streamwise vortices due to the imparted upwash/downwash effect.
    \item A secondary instability of the streaks of sub-harmonic sinuous type driven by the (1,3) mode induces spanwise meandering motions on the low-speed streaks --a process that is predominantly seeded by quadratic interactions between first and second  generation modes.
    \item In the late transitional stages prior to breakdown, coherent $\Lambda$-vortices appear and later disintegrate to small-scale structures, an event that is typically observed in canonical transitional boundary layer flows. The mean skin friction rises sharply towards the typical turbulent values.
\end{enumerate}

The present research demonstrated that it is sufficient to excite oblique waves via the first Mack instability mode to initiate the cascade of transitional mechanisms leading to turbulence in \ac{SWBLI}s. This statement stems from the fact that the optimal location and structure of the forcing waves are insensitive to the forcing amplitude. It is rather the non-linearity that, through a sequence of multi-modal interactions, paves the way to turbulence.

Future work following naturally from this study may involve the exploration of other ``optimal'' pathways to turbulence in stronger \ac{SWBLI}s (for example, where the incident shock angle $\theta$ is above criticality) displaying unstable self-excited instability mechanisms, such as the modal bubble resonance observed in many previous studies \cite{robinet2007bifurcations,cao_hao_klioutchnikov_wen_olivier_heufer_2022,hao_2023,song2023}. The capabilities of the framework may also be pushed beyond the transitional stages by incorporating closure models for the turbulent regime, at which point the number of collocation points necessary to resolve the turbulent scales would rise dramatically at the cost of simulation runtime and memory. In line with efforts to keep the cost of computations tractable, applications of this framework to flow control in \ac{SWBLI}s (heat load reduction, suppression of shock unsteadiness, transition control) are foreseen.

\begin{acknowledgments}
This work was funded by the Air Force Office of Scientific Research (AFOSR)/European Office of Aerospace Research and Development (EOARD) (Award FA8655-21-1-7009).
\end{acknowledgments}

\section*{Declaration of Interests}
The authors report no conflict of interest.

\section*{Author ORCIDs}
\orcidicon{0000-0002-2576-0685} F. Savarino \href{https://orcid.org/0000-0002-2576-0685}{https://orcid.org/0000-0002-2576-0685};

\orcidicon{0000-0002-2808-3886} D. Sipp \href{https://orcid.org/0000-0002-2808-3886}{https://orcid.org/0000-0002-2808-3886};

\orcidicon{0000-0001-6692-6437} G. Rigas \href{https://orcid.org/0000-0001-6692-6437}{https://orcid.org/0000-0001-6692-6437}.

\appendix
\section{\label{appA}Base-flow validation}

\begin{figure}
    \centering
    \includegraphics[width=0.5\textwidth]{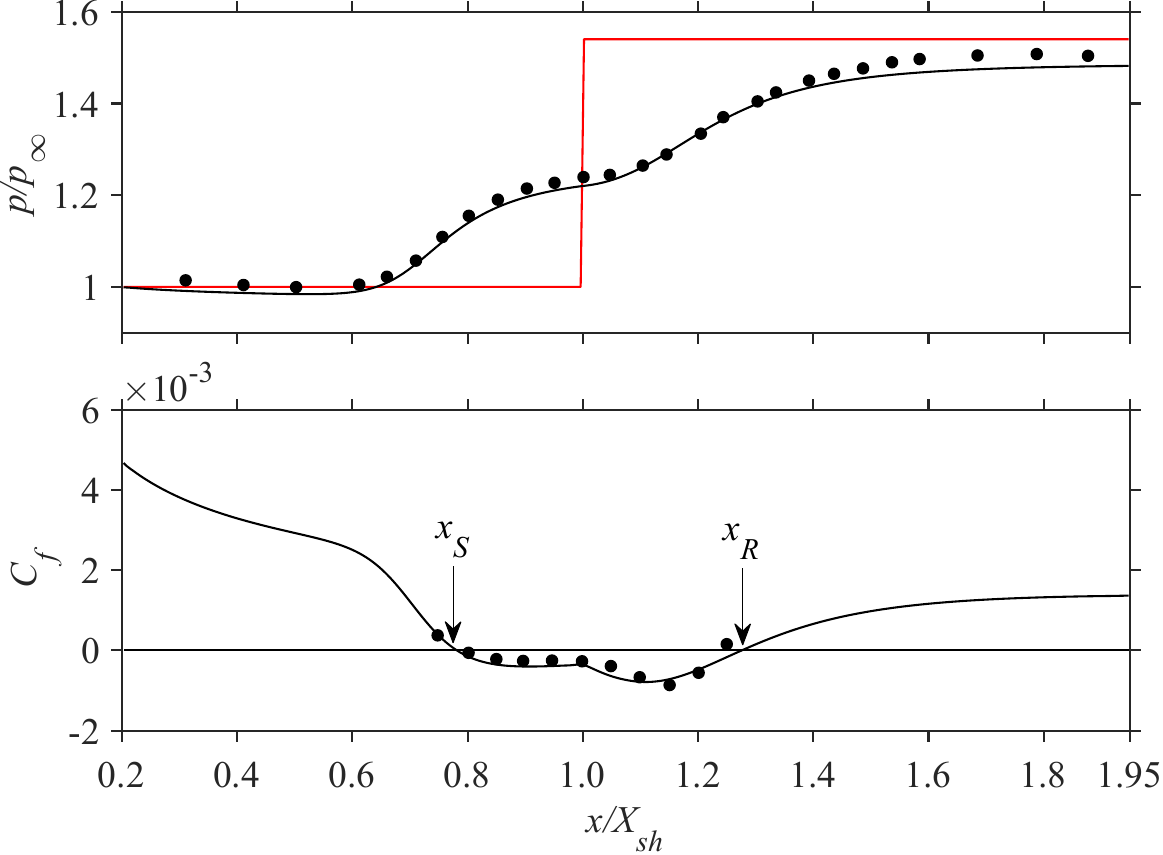}
    \caption{Validation of the base-flow of the $\theta=30.8^{\circ}$ \ac{SWBLI} compared to the experiment of \cite{degrez1987interaction}. (Top) wall pressure ratio from the numerical base-flow (black-solid), experimental data (black circles) and inviscid solution (red-solid). (Bottom) skin friction coefficient. Separation and reattachment points are indicated with the arrows. The peak reversed flow inside the separation bubble is $\max\{u_{\mathrm{min}}\}/U_{\infty}=5.01\%$.}
    \label{fig:baseflow validation}
\end{figure}

The calculation of the laminar base-flow for the \ac{SWBLI} is initialized with a Mach 2.15 \ac{ZPG} boundary layer solution on a sharp flat plate featuring a weak leading edge shock. The incident oblique shock is then introduced at angle $\theta$ relative to the streamwise direction and the interaction is converged to machine precision with a Newton iteration. The numerical base-flow for $\theta=30.8^{\circ}$ is shown in figure \ref{fig:baseflow validation}, where the wall pressure ratio (top) and skin friction coefficient (bottom) plots tell some important physical properties of the \ac{SWBLI}.

Before the shock impingement at $X_{sh}$, an \ac{APG} is felt by the boundary layer, named upstream influence \cite{babinsky_harvey_2011}. This leads to the first pressure rise, ultimately ascribable to a set of compression waves. The \ac{APG} is sufficiently strong to separate the boundary layer: a laminar separation point forms where $C_f$ first vanishes ($x_S$ in figure \ref{fig:baseflow validation}, bottom). At this point the flow splits into a separated shear layer and a region of reversed flow adjacent to the wall, where $C_f$ is negative. Over the first part of the separation zone up to $X_{sh}$, pressure is nearly constant, resulting in a plateau in the pressure ratio plot. The incident shock imposes the second pressure rise which is partly counteracted by the expansion waves emanating from the bubble's apex located approximately at $X_{sh}$. In the rear part of the separation zone, the re-accelerating shear layer gains sufficient momentum to fight the \ac{APG} and reattaches to the wall, forming a closed bubble. Weak shocklets at the reattachment zone deflect the flow back to parallel and the boundary layer slowly recovers its \ac{ZPG} features.

The comparison with \cite{degrez1987interaction}, who performed experiments at the same Mach number and shock angle, shows good agreement. Minimal discrepancies in the magnitude of the second pressure rise and in the location of the reattachment point are within acceptable bounds, since shock-induced separated shear layers are extremely sensitive to instabilities in experimental and real-life scenarios such that purely laminar separated interactions are impossible to reproduce outside the computational world.
\section{\label{appB}Linear stability analyses}

The linear stability of the non-linear dynamical system~\eqref{eq:state-space form} can be studied by linearizing the non-linear operator $\mathcal{N}(\mathbf{q})$ around the fixed-point of the system, also known as the base-state and denoted as $\mathbf{q}_0$. Let's consider the unforced system and Taylor expand the non-linear operator around $\mathbf{q}_0$, such that the evolution of the state can be written as
\begin{equation}
    \frac{\partial \mathbf{q}}{\partial t} = \mathcal{N}(\mathbf{q}_0) + \frac{\partial \mathcal{N}}{\partial \mathbf{q}}(\mathbf{q}_0) \left (\mathbf{q} - \mathbf{q}_0 \right ) + \frac{1}{2!} \frac{\partial^2 \mathcal{N}}{\partial \mathbf{q}^2}(\mathbf{q}_0) \left (\mathbf{q} - \mathbf{q}_0 \right )^2 + \mathcal{O}\left (\mathbf{q} - \mathbf{q}_0 \right )^3.
    \label{eq:Taylor expansion}
\end{equation}
Assuming the perturbation away from the base-state $\mathbf{q}'=\mathbf{q}-\mathbf{q}_0$ to be of infinitesimal amplitude $\left \| \mathbf{q}' \right \| \sim \varepsilon \ll 1$, we can truncate the expansion at the linear term, in which case we obtain two equations. At order 0, the equation for the base-state (or base-flow), 
\begin{equation}
    \mathcal{N}(\mathbf{q}_0) = \mathbf{0},
    \label{eq:base-flow}
\end{equation}
and at order 1, the equation for the evolution of linear perturbations around the base-flow,
\begin{equation}
    \frac{\partial \mathbf{q}'}{\partial t} = \frac{\partial \mathcal{N}}{\partial \mathbf{q}}(\mathbf{q}_0) \: \mathbf{q}',
    \label{eq:linear perturbations}
\end{equation}
where $\frac{\partial \mathcal{N}}{\partial \mathbf{q}}(\mathbf{q}_0) = \mathbf{J}$ is the Jacobian operator. The solution to eq.~\eqref{eq:base-flow}, $\mathbf{q}_0(x,y)$, is the 2D steady laminar solution of the governing \ac{N-S} equations (eq.~\eqref{eq:governing equations}), while eq.~\eqref{eq:linear perturbations} can be studied to detect any modal instabilities of the linearized system dynamics.

\subsection{\label{appB:global EVP}Global eigenvalue problem}
We introduce the Fourier ansatz for the linear perturbations,
\begin{equation}
    \mathbf{q}'(x,y,z,t) = \hat{\mathbf{q}}(x,y) \: \exp{\left ( \text{i}\beta z - \lambda t \right )} + \text{c.c.},
    \label{eq:global modes}
\end{equation}
where $\beta$ is the real spanwise wavenumber and $\lambda=\sigma+\text{i}\omega$ is the complex eigenvalue associated with the spatial eigenmode $\hat{\mathbf{q}}(x,y)$. Injecting eq.~\eqref{eq:global modes} into eq.~\eqref{eq:linear perturbations}, we derive the global eigenvalue problem, 
\begin{equation}
    - \lambda \hat{\mathbf{q}} = \mathbf{J} \hat{\mathbf{q}},
    \label{eq:global eigenvalue problem}
\end{equation}
whose solutions are the eigenpairs $\left \{ \lambda, \hat{\mathbf{q}} \right \}_i$. If one of the eigenvalues has negative real part, i.e. $\sigma<0$, the flow is globally unstable and the global eigenmode grows exponentially in time. On the other hand, stable eigenmodes decay exponentially to zero. The imaginary part of the eigenvalue, $\omega$, is the pulsation. If zero, the corresponding eigenmode is non-oscillatory. Through the real wavenumber $\beta$ we can study eigenmodes that are periodic in the spanwise direction $z$.

\begin{figure}
    \centering
    \includegraphics[width=\textwidth]{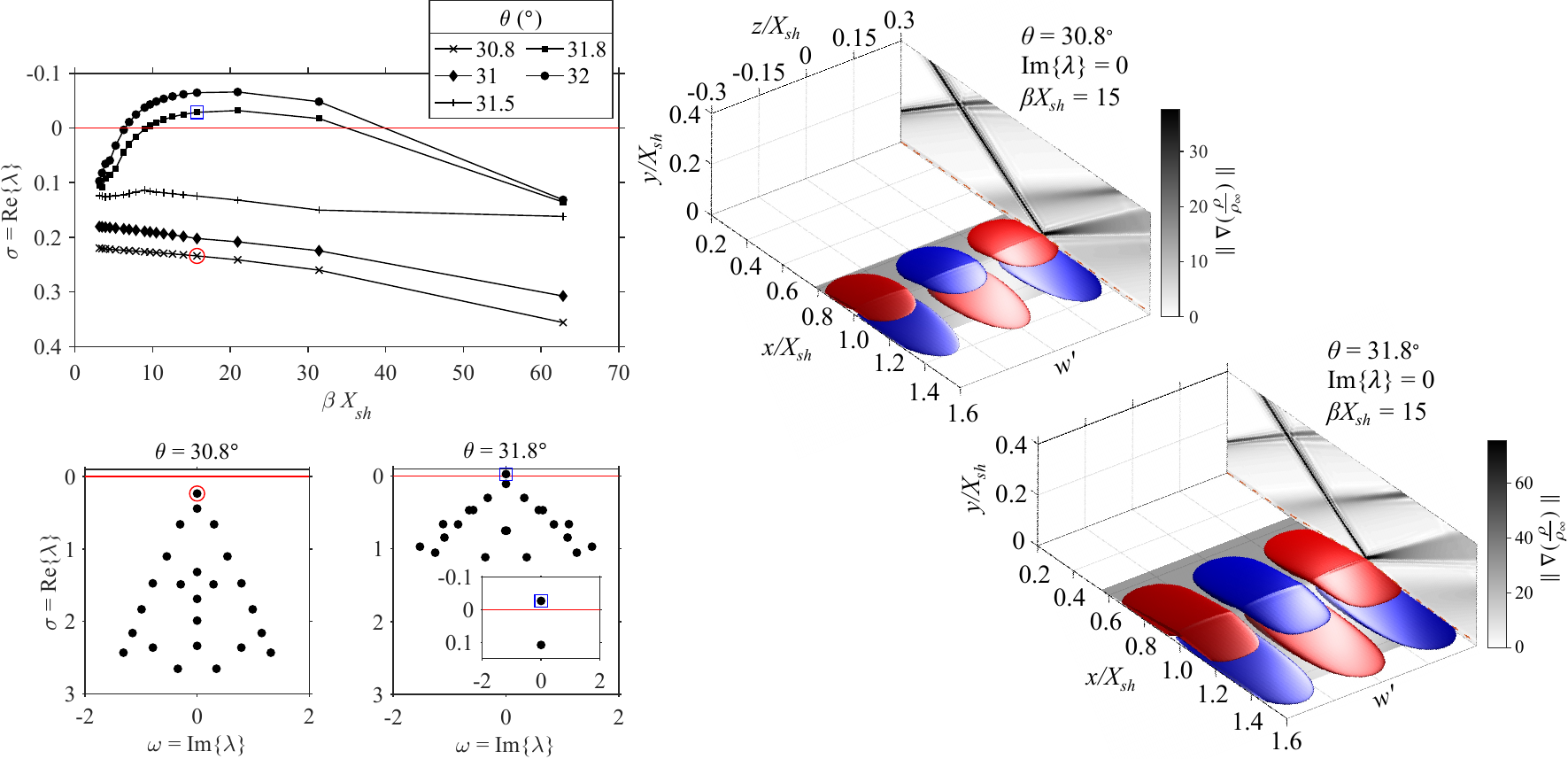}
    \caption{Bubble breathing instability mode computed from the global stability eigen-problem on the \ac{SWBLI} base-flow. (Top left) temporal amplification factor ($\sigma=\mathrm{Re}\{\lambda\}$) of the least stable/most unstable eigenvalue of the eigenspectrum computed for a range of spanwise wavenumbers and for five \ac{SWBLI}s of different strength. (Bottom left) eigenspectra of a globally stable ($\theta=30.8^{\circ}$, $\sigma_{1}>0$) and unstable ($\theta=31.8^{\circ}$, $\sigma_{1}<0$) base-flow at $\beta X_{sh}=15$. (Right) eigenmode shape for the globally (top) stable and (bottom) unstable cases. The separation bubble and magnitude of the 1st density gradient are also shown.}
    \label{fig:eigenmodes}
\end{figure}



In the top left panel of figure \ref{fig:eigenmodes} we plot the amplification rate of the least stable eigenvalue from five base-flows over a range of spanwise wavenumbers $\beta X_{sh}$. It is observed that the \ac{SWBLI} base-flow is globally unstable only for a selective range of spanwise wavenumbers and beyond a critical shock angle $\theta$, which in our case is in the range $31.5^{\circ}-31.8^{\circ}$. This indicates there is a 3D unstable eigenmode that is generated intrinsically by the flow and the dependency on the shock angle $\theta$ suggests it is related to the size of the separation bubble, which increases with $\theta$. In the two bottom left panels we show the eigenvalue spectra of a globally stable ($\theta=30.8^{\circ}$) and unstable ($\theta=31.8^{\circ}$) case at $\beta X_{sh}=15$ and notice that the unstable mode is non-oscillatory (zero imaginary part). These features agree with the existing literature \cite{robinet2007bifurcations,ganapathisubramani2009low,clemens2014low,hao_2023} reporting a global instability mechanism over a number of different \ac{SWBLI} geometries and configurations. On the right panels, the least stable (for $\theta=30.8^{\circ}$) and unstable (for $\theta=31.8^{\circ}$) eigenmodes from the $\beta X_{sh}=15$ eigenspectra are plotted. The modes display the same 3D structure with characteristic spanwise wavelength $\lambda_z\sim L_{\mathrm{sep}}$ and share the same physics, except that one exponentially decays with time and the other gets self-amplified via a global resonance. The spatial arrangement of spanwise velocity disturbances located in the separation zone illustrates the physical mechanism of the instability which, upon non-linear saturation, creates 3D spanwise corrugations of the separation bubble \cite{guiho_alizard_robinet_2016,cao_hao_klioutchnikov_wen_olivier_heufer_2022,hao_2023,song2023}.

\subsection{\label{appB:resolvent analysis}Resolvent analysis}
Even in globally stable fluid systems such as boundary layers, disturbances can grow in time due to the non-normality of the linearized \ac{N-S} operator, in particular, stemming from the shear in the flow. Flows of this kind are called amplifier flows \cite{huerre1990local}. In such cases, all the eigenvalues of the Jacobian operator decay in time, however the non-normality makes it possible for some eigendirections to be almost aligned and give rise to transient growth mechanisms \cite{schmid_henningson2001stability}. Therefore, it is useful to look into the pseudo-resonances of globally stable flows to external disturbances through the resolvent operator.

For this analysis we consider the linearized system~\eqref{eq:linear perturbations} but we introduce a small amplitude external forcing term $\mathbf{f}'(x,y,z,t) = \hat{\mathbf{f}}(x,y) \: \exp{\left [\text{i}(\beta z + \omega t) \right ]} + \text{c.c.}$ on the right-hand side to account for external disturbances which may be amplified by the linearized flow dynamics. Clearly, the response takes a similar form $\mathbf{q}'(x,y,z,t) = \hat{\mathbf{q}}(x,y) \: \exp{\left [\text{i}(\beta z + \omega t) \right ]} + \text{c.c.}$. Injecting these expressions in the forced linear system yields 
\begin{equation}
    \hat{\mathbf{q}} = \left ( \text{i} \omega \mathbf{I} - \mathbf{J}(\omega,\beta) \right )^{-1} \hat{\mathbf{f}},
    \label{eq:linear resolvent}
\end{equation}
which is the input/output relation between the forcing (input) and the response (output) of the system. The resolvent operator $\mathcal{H} = \left( \text{i} \omega \mathbf{I} - \mathbf{J}(\omega,\beta) \right)^{-1}$ is the transfer function relating the two. The Jacobian $\mathbf{J}(\omega,\beta)$ takes into account both 2D and 3D, steady and unsteady disturbances with characteristic frequency $\omega$ and spanwise wavenumber $\beta$. After spatial discretization, the linear resolvent formulation can be rewritten in compact form as
\begin{equation}
    \hat{\mathbf{q}} = \mathbf{H}\mathbf{MP}\hat{\mathbf{f}}.
    \label{eq:linear resolvent discrete}
\end{equation}
where $\mathbf{M}$ is the mass matrix and $\mathbf{H}$ is the discretized resolvent. The forcing may be applied only to specific components and/or regions of the flow domain through the restriction/prolongation matrix $\mathbf{P}$. The optimal forcing/response modes are computed by optimizing the energy gain,
\begin{equation}
    \sigma^2 = \max_{\hat{\mathbf{f}}} \frac{\| \hat{\mathbf{q}} \|^2_E}{\| \hat{\mathbf{f}} \|^2_F},
    \label{eq:resolvent gain}    
\end{equation}
with $\|\cdot\|_E$ and $\|\cdot\|_F$ the user-selected norms to evaluate the amplitude of the response and forcing, respectively. These are:
\begin{equation}
    \| \hat{\mathbf{q}} \|^2_E = \hat{\mathbf{q}}^* \mathbf{Q}_E \hat{\mathbf{q}}, \:\:
    \| \hat{\mathbf{f}} \|^2_F = \hat{\mathbf{f}}^* \mathbf{Q}_F \hat{\mathbf{f}}.
    \label{eq:energy norms}
\end{equation}
For compressible flows, the energy of the response is defined by $\mathbf{Q}_E = \mathbf{Q}_{\text{Chu}}$ from Chu's energy \cite{george2011chu} to account for the pressure ($\hat{p}$) and entropy ($\hat{s}$) disturbances,
\begin{equation}
    E_\text{Chu} = \hat{\mathbf{q}}^* \mathbf{Q}_{\text{Chu}} \hat{\mathbf{q}} = \frac{1}{2} \int_\Omega \left( \rho_0|\hat{\mathbf{u}}|^2 + \frac{1}{\gamma p_0} \hat{p}^2 + \gamma (\gamma - 1) M^4\, p_0\, \hat{s}^2 \right) \text{d}\Omega. 
    \label{eq:Chu energy}    
\end{equation}
The energy of the forcing is computed simply from the $L_2$-norm. For more details refer to \cite{hanifi1996transient,george2011chu}. 

We solve the optimization in eq.~\eqref{eq:resolvent gain} by formulating a generalized eigenvalue problem, whose solutions are the optimal forcing $\hat{\mathbf{f}}$ (eigenvector) and the squared gain $\sigma^2$ (eigenvalue). We repeat this calculation over a range of frequencies and wavenumbers in order to identify selective frequency-wavenumber ranges for disturbance amplification. Through eq.~\eqref{eq:linear resolvent discrete} we directly find the corresponding system's optimal response. The forcing is assumed to excite the system's dynamics continuously in order for the response not to decay to zero.

\begin{figure}
    \centering
    \includegraphics[width=0.4\textwidth]{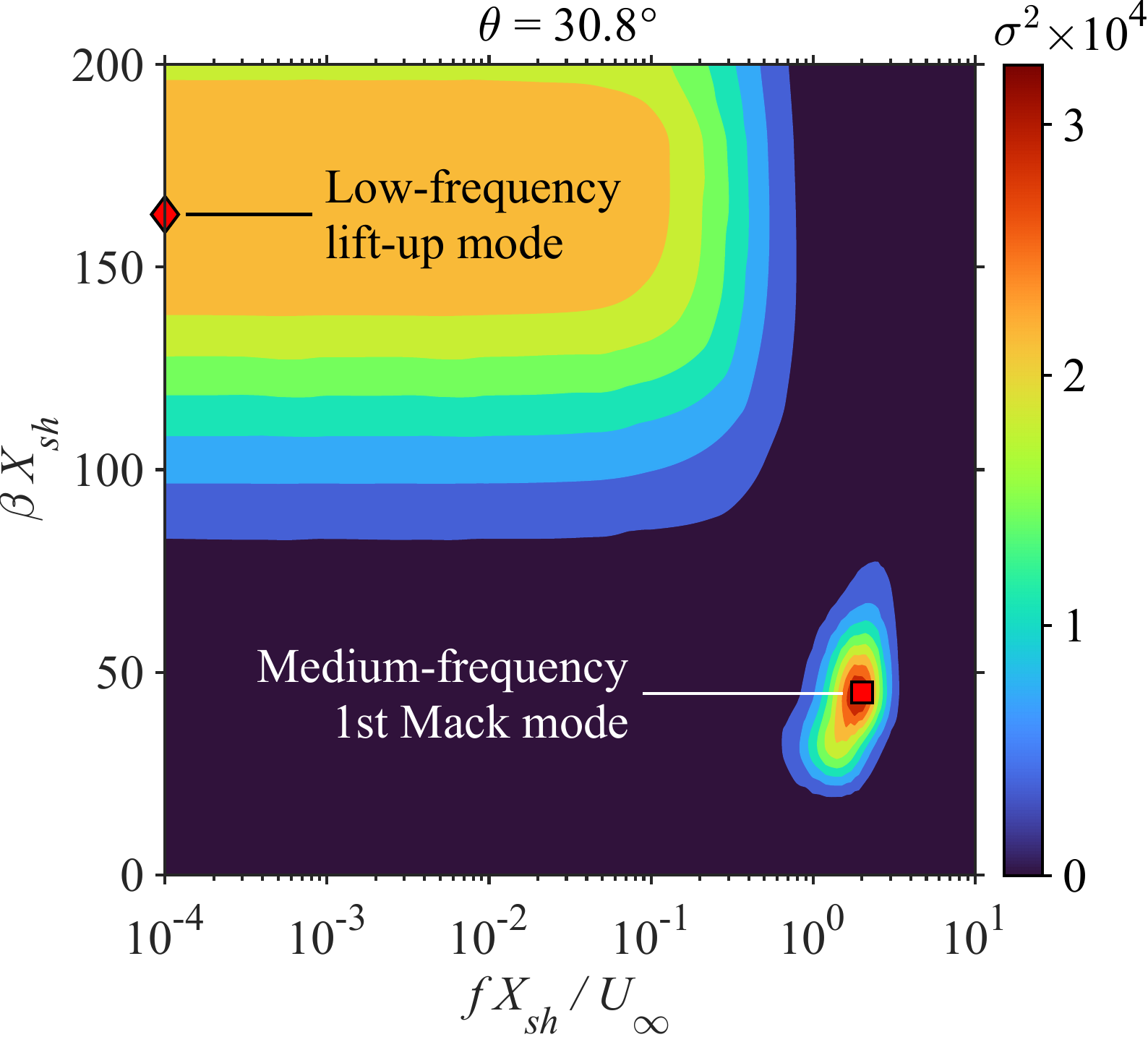}
    \caption{Contours of the (squared) resolvent gain spectrum in the frequency-spanwise wavenumber space for the $\theta=30.8^{\circ}$ \ac{SWBLI}. The leading singular value is shown. Local peaks in the spectrum are annotated with a square/diamond marker referring to the first/second most unstable resolvent mode.}
    \label{fig:resolvent gain}
\end{figure}

In figure \ref{fig:resolvent gain} the resolvent spectrum for the $\theta=30.8^{\circ}$ \ac{SWBLI} reveals two regions of high amplification. The maximum gain is recorded at $(fX_{sh}/U_{\infty},\beta X_{sh})=(2,45)$ which sits in the medium-frequency range, while a second local maximum is found at nominally zero frequency and $\beta X_{sh}=163$ within the low-frequency plateau (the frequency axis in figure \ref{fig:resolvent gain} does not go to zero, as log-scale is used). 

\begin{figure}
    \centering
    \includegraphics[width=\textwidth]{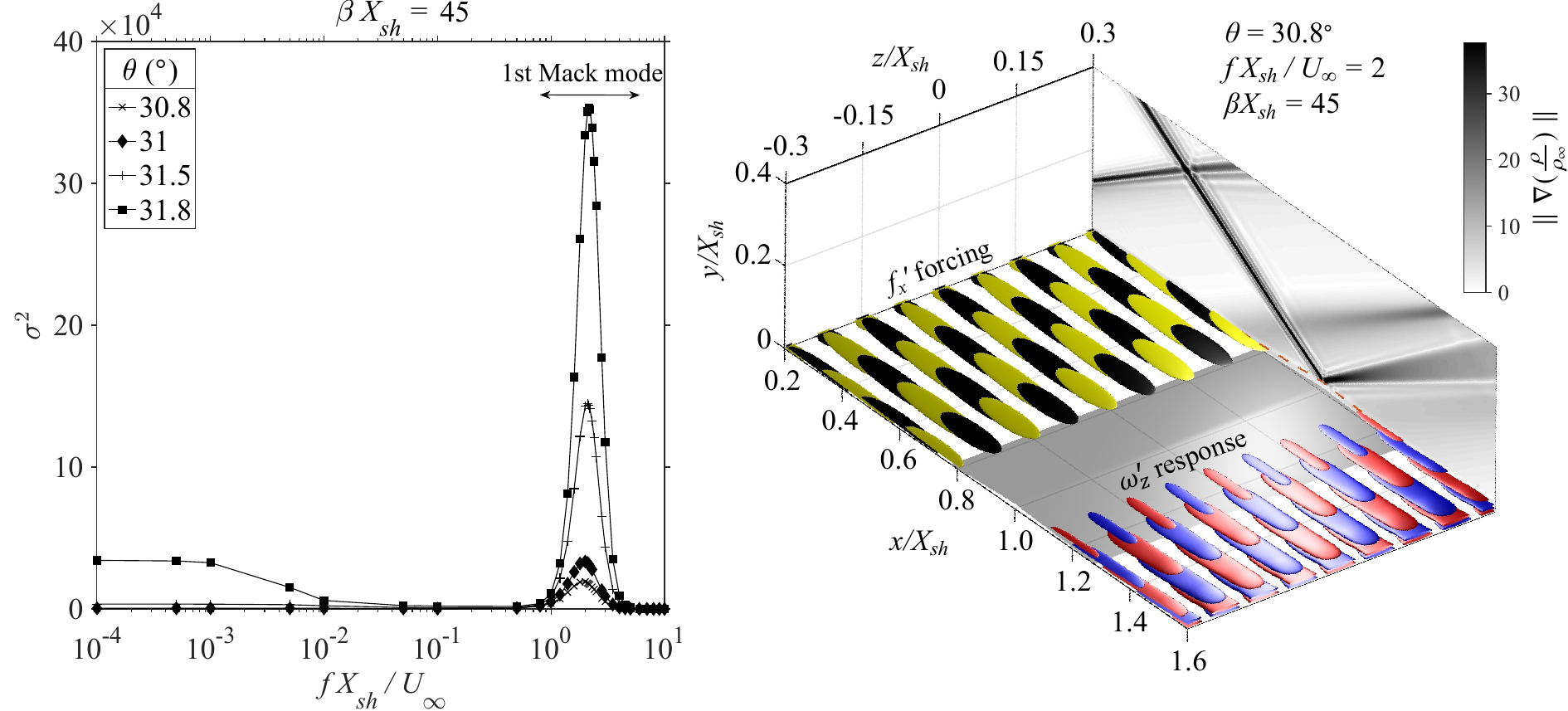}
    \caption{Medium-frequency linear instability: 1st Mack mode. (Left) squared resolvent gain distribution for a range of temporal frequencies and four \ac{SWBLI}s of different strength at $\beta X_{sh}=45$. The leading singular value is shown. (Right) 3D isosurfaces (yellow/red: positive, black/blue: negative) of the most unstable resolvent forcing/response modes of the $\theta=30.8^{\circ}$ \ac{SWBLI} at $(fX_{sh}/U_{\infty},\beta X_{sh})=(2,45)$. The separation bubble and magnitude of the 1st density gradient are also shown.}
    \label{fig:mack mode}
\end{figure}

The medium-frequency peak, highlighted in figure \ref{fig:mack mode} (left) for a number of interaction strengths, denotes the 1st Mack instability mode --the 3D (oblique) version being more unstable than the 2D (planar) one \cite{bugeat_robinet_chassaing_sagaut_2022}. On the right panel, the corresponding forcing/response resolvent modes display wave packets in checkerboard arrangement with spanwise wavelength $\lambda_z\approx 5\delta^{*}$. The forcing is located upstream of separation and the waves are oriented against the mean shear direction, while the response waves grow in the separated shear. This shear phenomenon is responsible for the vortex shedding dynamics in the medium-frequency band \cite{guiho_alizard_robinet_2016,bugeat_robinet_chassaing_sagaut_2022}.

\begin{figure}
    \centering
    \includegraphics[width=\textwidth]{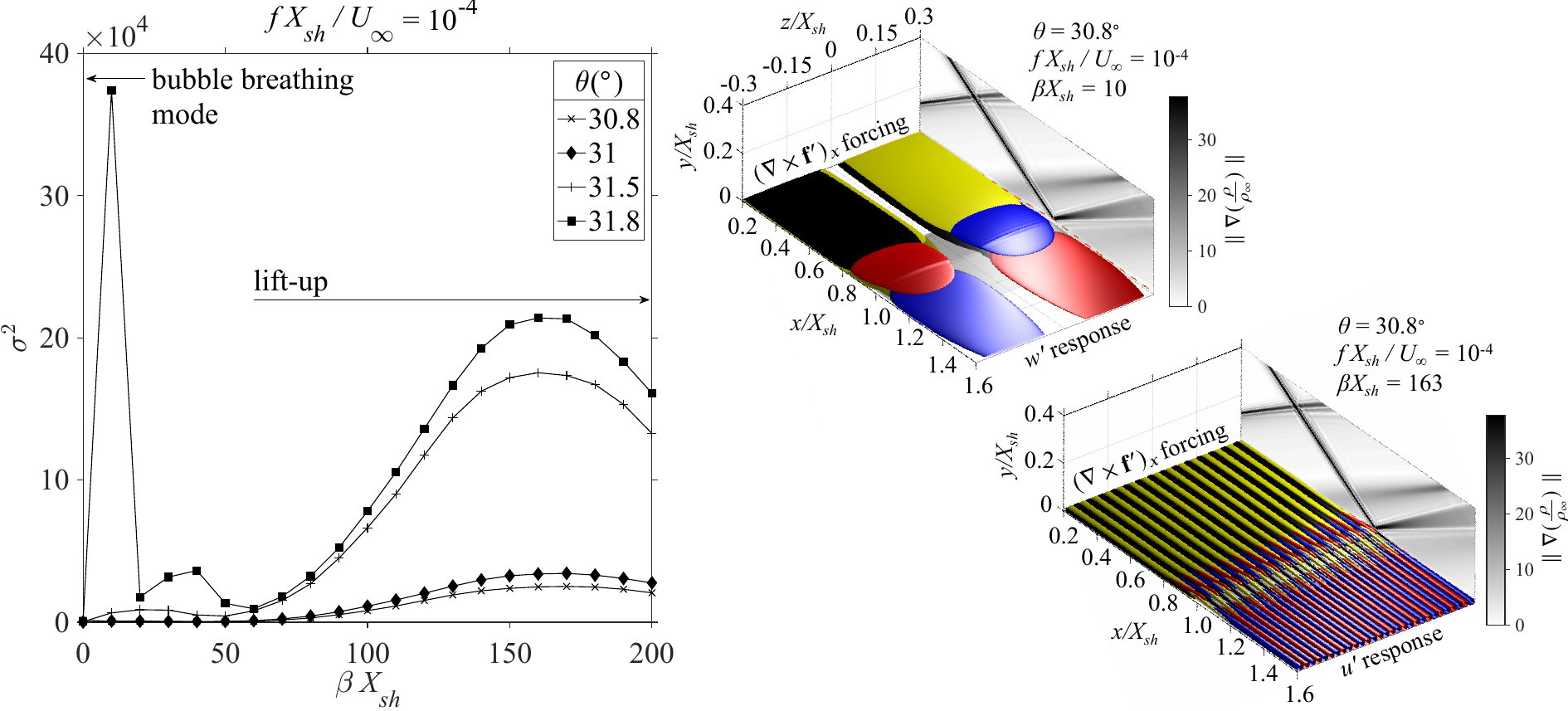}
    \caption{Low-frequency linear instabilities: bubble breathing and lift-up modes. (Left) squared resolvent gain distribution for a range of spanwise wavenumbers and four \ac{SWBLI}s of different strength at $fX_{sh}/U_{\infty}=10^{-4}$. The leading singular value is shown. (Right) 3D resolvent forcing/response isosurfaces (yellow/red: positive, black/blue: negative) of the (top) globally stable bubble breathing mode at $(fX_{sh}/U_{\infty},\beta X_{sh})=(10^{-4},10)$ and (bottom) lift-up mode at $(fX_{sh}/U_{\infty},\beta X_{sh})=(10^{-4},163)$ for the $\theta=30.8^{\circ}$ \ac{SWBLI}. The separation bubble and magnitude of the 1st density gradient are also shown.}
    \label{fig:lift up and bubble breathing modes}
\end{figure}

The low-frequency band of the spectrum in figure \ref{fig:lift up and bubble breathing modes} (left) detects two different disturbance mechanisms. A strongly 3D mode ($\lambda_z \sim \delta^{*}$) ascribed to non-modal, transient growth of streamwise streaks via the lift-up mechanism \cite{klebanoff_tidstrom_sargent_1962,hanifi1996transient,schmid_henningson2001stability,andersson_brandt_bottaro_henningson_2001} and a weakly 3D mode (spanwise wavenumber one order of magnitude lower than that of streaks) that displays the features of a global resonance, i.e. sudden spike in the gain value for the super-critical $\theta=31.8^{\circ}$ \ac{SWBLI}. On the right panels, the spatial structures of these two modes for the $\theta=30.8^{\circ}$ interaction emphasize the differences in these mechanisms. The weakly 3D response mode is strikingly similar to the eigenmode in figure \ref{fig:eigenmodes}, confirming the system's pseudo-resonances captured by the resolvent match the true resonances, should they be present \cite{Symon2018nonnormality_resolvent}. Furthermore, the computed forcing shows that this instability is optimally triggered by vortical excitations upstream of the separation shock foot. While this mode is nominally zero-frequency/non-oscillatory (figure \ref{fig:eigenmodes}), any small fluctuations in both high-fidelity numerical simulations and experiments can trigger a low-frequency mechanism that couples the separation bubble mode with the separation shock, resulting in a breathing motion of the bubble and a back and forth oscillation of the separation shock \cite{ganapathisubramani2009low,clemens2014low}.

Overall, our linear stability analyses agree with the broader literature reviewed by \cite{gaitonde2015progress}, according to which the fundamental \ac{SWBLI} dynamics can be broken down into three main frequency ranges: 
\begin{enumerate}
    \item Low-frequency ($St_{L_{\mathrm{sep}}} \sim 10^{-4}$-$10^{-2}$): bubble breathing mode and streaks,
    \item Medium-frequency ($St_{L_{\mathrm{sep}}} \sim 10^{-1}$-$10^{0}$): shear layer modes (Mack, Kelvin-Helmholtz instabilities), and
    \item High-frequency ($St_{L_{\mathrm{sep}}} \sim 10^{1}$): free-stream turbulent fluctuations (only in turbulent \ac{SWBLI}s).
\end{enumerate}
\section{\label{appC}Parametric studies}

\subsection{\label{appC:super-harmonic forcing}Super-harmonic forcing}
We explore the super-harmonic forcing configuration to see if the route to turbulence of the $\theta=30.8^{\circ}$ \ac{SWBLI} is affected. The set-up is the same as in \S\ref{sec:optimal non-linear mechanisms for laminar--turbulent transition}, but we also extend the optimization to the forcing harmonics $(\pm2\omega,0\beta)$, $(\pm2\omega,\pm1\beta)$, $(\pm1\omega,\pm2\beta)$, $(0\omega,\pm2\beta)$ and $(\pm2\omega,\pm2\beta)$.

\begin{figure}
    \centering
    \includegraphics[width=\textwidth]{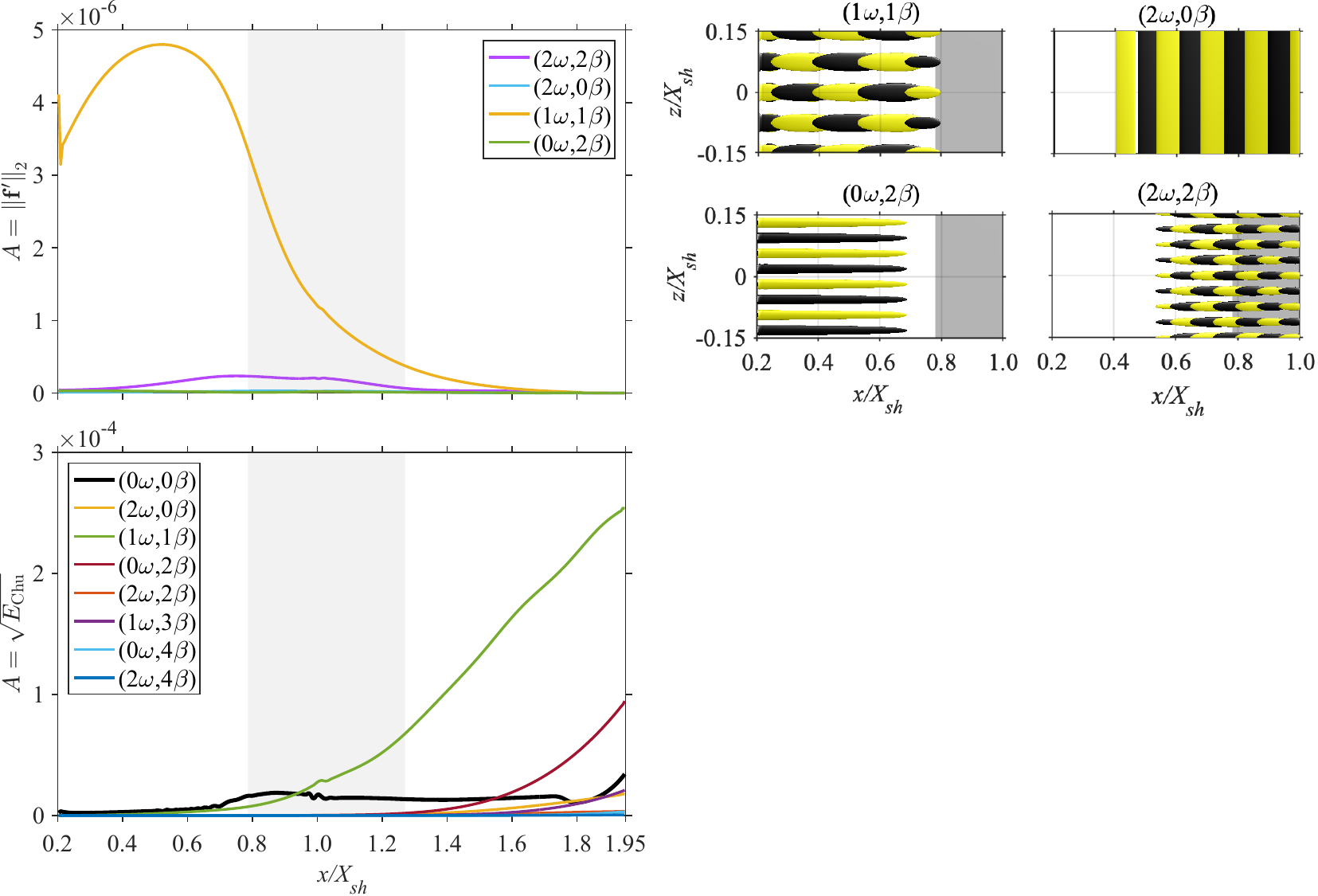}
    \caption{Optimal forcing/response solution at low amplitude $A=0.5\times10^{-5}$ computed from the $N=2,M=4$ system with the super-harmonic forcing configuration. Amplitudes of (top left) forcing harmonics based on the $L_2$-norm and (bottom left) response harmonics based on Chu's energy definition \cite{george2011chu}. The mean separation length is plotted in gray. (Right) 3D structure of the forcing harmonics using (yellow: positive, black: negative) isosurfaces of $(\nabla \times \boldsymbol{f}')_{z}$ for the $(1\omega,1\beta)$, $(2\omega,0\beta)$ and $(2\omega,2\beta)$ harmonics and $(\nabla \times \boldsymbol{f}')_{x}$ for the $(0\omega,2\beta)$ harmonic. The time-averaged separation bubble is superimposed with a gray isosurface.}
    \label{fig:forcing/response solution low amplitude super-harmonic}
\end{figure}

In figure \ref{fig:forcing/response solution low amplitude super-harmonic} we plot the solution at low amplitude $A=0.5\times10^{-5}$ and observe that the fundamental $(1,1)$ forcing mode makes $\approx93\%$ of the total forcing, while super-harmonic planar $(2,0)$ and oblique $(2,2)$ wave-type forcings and steady streak-type $(0,2)$ forcing only 7\%. This means the optimal forcing is the fundamental oblique $(1,1)$ mode for which reason the response of the flow is virtually identical to the fundamental forcing case in figure \ref{fig:forcing/response solution low amplitude}.

\subsection{\label{appC:frequency scan}Frequency-spanwise wavenumber scan}
The analysis performed in \S\ref{sec:optimal non-linear mechanisms for laminar--turbulent transition} assumes that forcing the flow at the most unstable frequency-spanwise wavenumber of the linear 1st Mack mode yields maximum increase in the drag. Because the non-linear behavior of the fluid system is initially governed by a substantially linear primary instability mechanism up to the reattachment zone, where the non-linearity of the system starts to play a role, the assumption is indeed fair. Nonetheless, there might be other frequencies that could result in more efficient pathways to turbulence. The first reason is that the cost function of the optimization~\eqref{eq:cost function} is different to that of linear resolvent analysis~\eqref{eq:resolvent gain}. Secondly, the optimal frequency-spanwise wavenumber pair may well be affected by the non-linear behavior of the system. These reasons motivate the parametric study we perform on the frequency-spanwise wavenumber plane.

\begin{figure}
    \centering
    \includegraphics[width=0.4\textwidth]{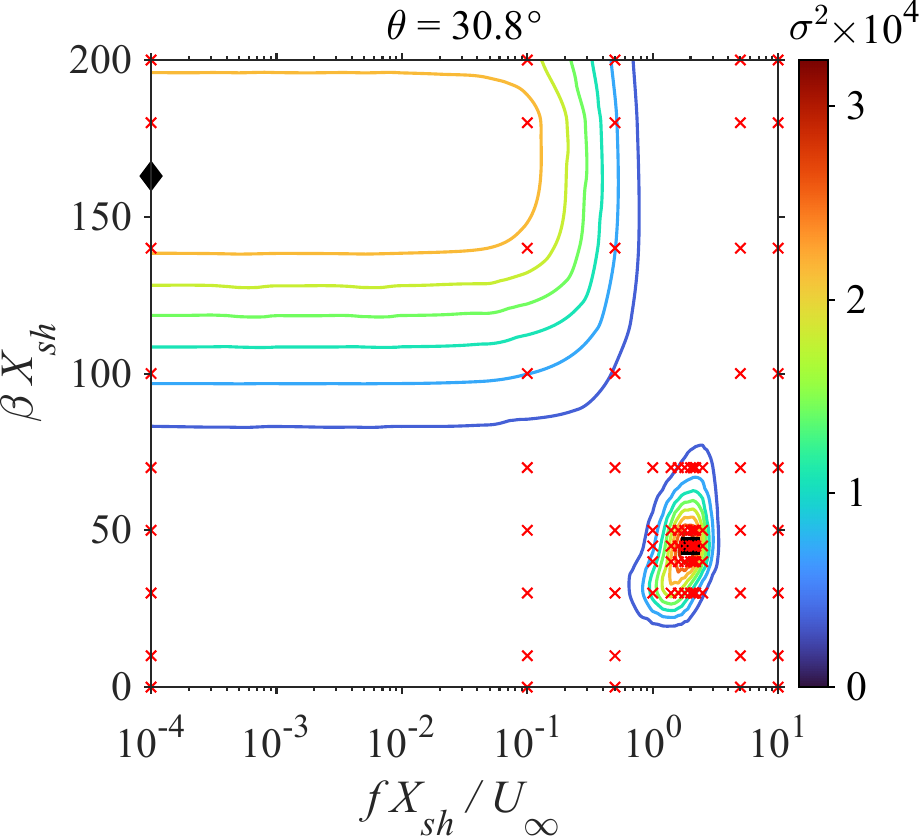}
    \caption{Parametric study on the frequency-spanwise wavenumber plane. The test cases are marked with red crosses on underlying contours of the linear resolvent gain, identical to figure \ref{fig:resolvent gain}. In total, 99 non-linear optimization cases are computed with system $N=2,M=4$.}
    \label{fig:map scan}
\end{figure}

As shown in figure \ref{fig:map scan}, we seed the plane non-uniformly. We scan the neighborhood of the 1st Mack mode more densely to capture any changes, if any, in the optimal frequency, while the high gain plateau of streak-type instabilities is coarsely seeded given the slow rate of change of the gain. This non-uniform distribution of test cases allows to bound the number of optimizations to 99 and therefore keep the cost of the parametric study feasible. We employ system $N=2,M=4$ and perform the amplitude continuation up to $A=2.4\times10^{-5}$ for each test case. Furthermore, to alleviate the computational cost, we choose the fundamental forcing configuration for this study.

\begin{figure}
    \centering
    \includegraphics[width=0.8\textwidth]{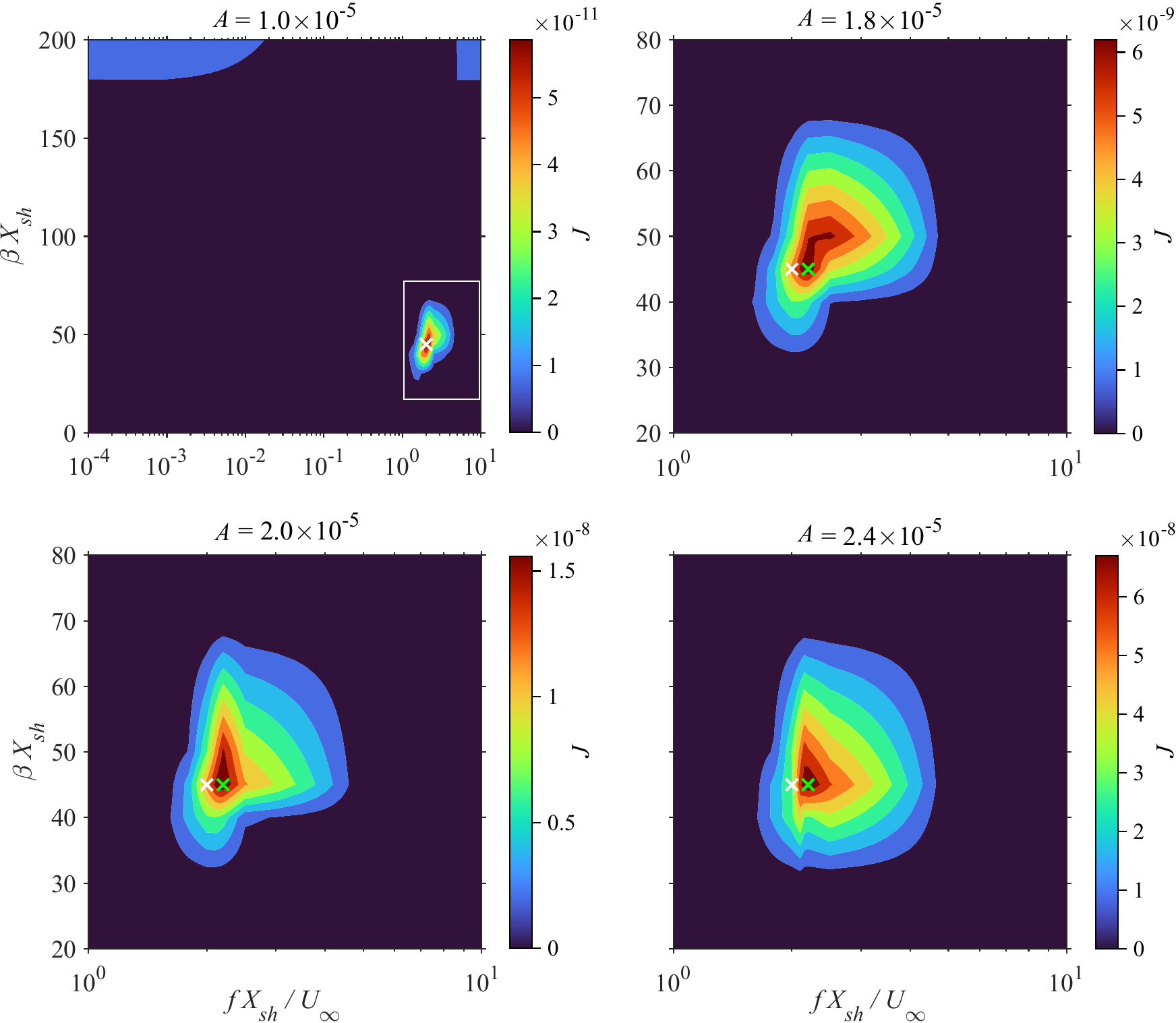}
    \caption{Contours of the cost function $J$ computed on the frequency-spanwise wavenumber plane for the test cases defined in figure \ref{fig:map scan}. Four forcing amplitudes are shown from $1.0\times10^{-5}$ to $2.4\times10^{-5}$. The white cross indicates the most unstable linear mode at $(fX_{sh}/U_{\infty},\beta X_{sh})=(2,45)$. The green cross is the non-linear counterpart at $(fX_{sh}/U_{\infty},\beta X_{sh})=(2.2,45)$. The white box in the $A=1.0\times10^{-5}$ map marks the section of the plane shown for the other amplitudes.}
    \label{fig:cost map scan}
\end{figure}

In figure \ref{fig:cost map scan} we show the cost function $J$ in eq.~\eqref{eq:cost function} on the frequency-spanwise wavenumber plane for four forcing amplitudes. From the low amplitude solution at $A=1.0\times10^{-5}$ we observe that the medium-frequency range around the optimal linear 1st Mack mode leads to the highest values of the cost function, whereas other areas of the map display substantially lower values. This result clearly demonstrates the dominance of the 1st Mack mode as the primary instability in the \ac{SWBLI} flow.

\begin{figure}
    \centering
    \includegraphics[width=\textwidth]{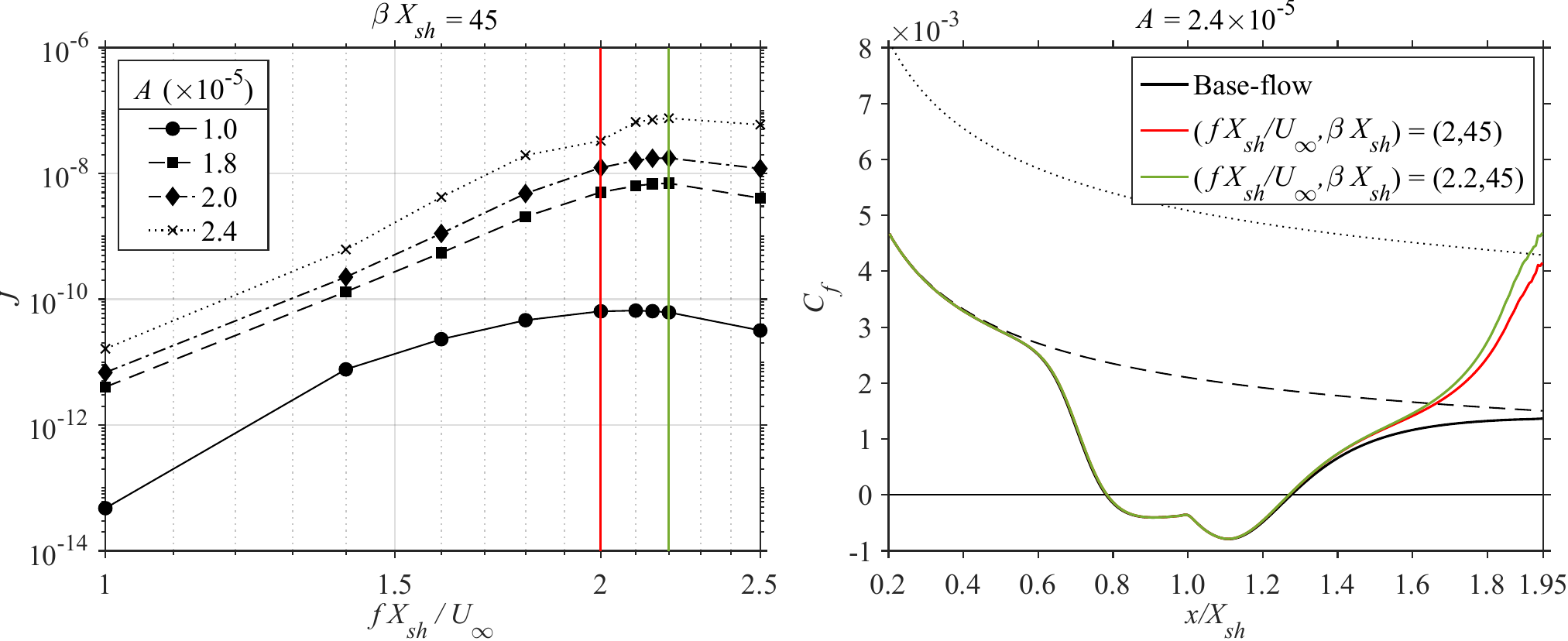}
    \caption{(Left) cost function $J$ versus frequency at $\beta X_{sh}=45$ for four forcing amplitudes. The red and green lines mark the optimal frequencies computed from linear and non-linear analyses, respectively. (Right) comparison of the mean skin friction coefficient at these two frequencies for amplitude $A=2.4\times10^{-5}$. Black-dashed line: laminar \ac{ZPG} boundary layer \cite{Howarth_1948}. Black-dotted line: turbulent \ac{ZPG} boundary layer \cite{Franko_Lele_2013}. Black-solid: laminar \ac{SWBLI} base-flow.}
    \label{fig:cost and Cf linear vs nonlinear}
\end{figure}

If we inspect more closely the medium-frequency range of the Mack mode for the higher amplitude cases $A=1.8$-$2.4\:\times10^{-5}$, we notice a small shift in the optimal frequency, but not in the spanwise wavenumber. This shift is visible more clearly in figure \ref{fig:cost and Cf linear vs nonlinear} (left) where we extract the cost function for the four amplitude cases over the range $1\leq fX_{sh}/U_{\infty}\leq2.5$. The frequency shift is only +0.2 and forcing the flow at this new frequency slightly increases the skin friction coefficient --see figure \ref{fig:cost and Cf linear vs nonlinear} (right). Anyhow, the instability mechanisms at play are identical to \S\ref{sec:optimal non-linear mechanisms for laminar--turbulent transition}.
\section{\label{appD}Görtler analysis}

Flows with streamline curvature may host centrifugal-type instabilities \cite{Gortler1941,floryan_saric_1982,hall_malik_1989,floryan1991,Saric1994}. These manifest as secondary flows that interact with the main stream and participate in the transition process \cite{roghelia_olivier_egorov_chuvakhov_2017}. Useful parameters to detect sites of centrifugal instability amplification are the Görtler number \cite{Gortler1941,smiths_dussauge_2006,currao_2020} and the curvature parameter \cite{floryan1991}. In our analysis we consider the Görtler number definition from \cite{smiths_dussauge_2006} for a compressible turbulent boundary layer, 
\begin{equation}
    G_T = \frac{\theta}{0.018\delta^{*}}\sqrt{\frac{\theta}{\mathcal{R}}},
    \label{eq:Görtler number}
\end{equation}
that allows convenient comparison with the criterion of \cite{Gortler1941} for which $G_T\geq0.6$ indicates unstable regions for Görtler mode amplification. 
\begin{figure}
    \centering
    \includegraphics[width=0.8\textwidth]{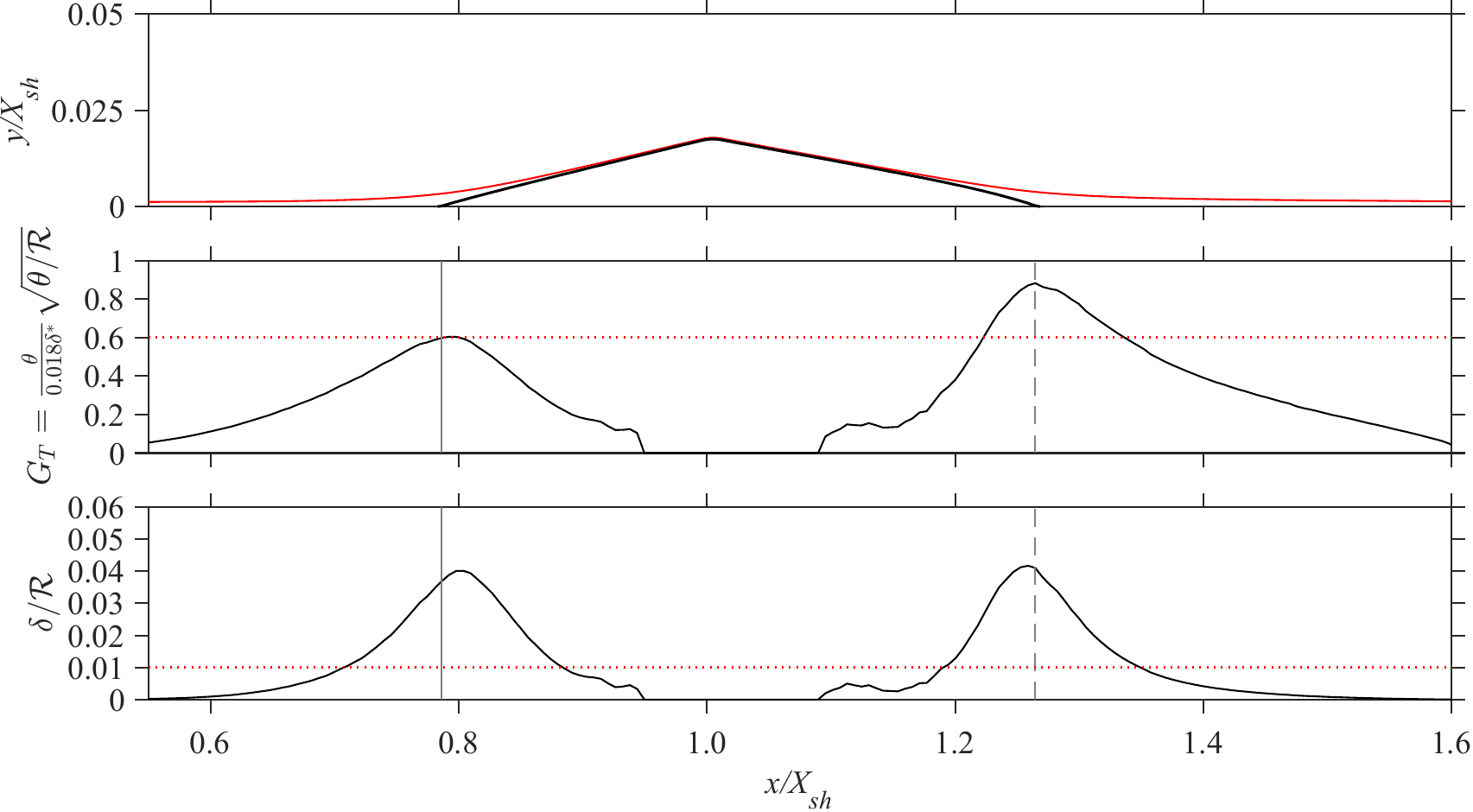}
    \caption{Görtler analysis for an open streamline obtained from the $\theta=30.8^{\circ}$ \ac{SWBLI} mean-flow at $A=0.5\times10^{-5}$. (Top) dividing streamline of the mean separation bubble (black-solid line) and open streamline (red-solid line). (Middle) Görtler number definition from \cite{smiths_dussauge_2006} and critical value 0.6 (red-dotted line) from \cite{Gortler1941}. (Bottom) curvature parameter and critical value 0.01 (red-dotted line) from \cite{floryan1991}.}
    \label{fig:Görtler analysis}
\end{figure}
The integral boundary layer parameters $\delta^{*}$ and $\theta$, and the radius of curvature $\mathcal{R}$ are computed from an open streamline of the time- and spanwise-averaged \ac{SWBLI} flow at low amplitude $A=0.5\times10^{-5}$. The chosen streamline is adjacent to the separation bubble, as shown in figure \ref{fig:Görtler analysis} (top), and is a typical choice to evaluate the curvature of the flow in both oblique/reflected shock \cite{kontis_erdem_johnstone_murray_steelant_2013,Giepman_Schrijer_vanOudheusden_2018,Pasquariello_Hickel_Adams_2017,currao_2021,jiao_ma_xue_wang_chen_cheng_2024} and compression ramp/cone-flare \cite{loginov_adams_zheltovodov_2006,kuehl_paredes_2016,roghelia_olivier_egorov_chuvakhov_2017,currao_2020,zhao_ma_chen_zhang_hao_wen_2024,dixit_kumar_vadlamani_tsuboi_2025,sun_yu_li_zhang_2025} configurations. As the Görtler number (middle) and curvature parameter (bottom) values show, the separation and reattachment are critical sites of the shear layer where centrifugal effects can manifest. In particular, the reattachment zone displays the highest Görtler number ($G_T\approx0.9$) and curvature parameter ($\delta/\mathcal{R}\approx0.04$), which satisfy the criteria \cite{Gortler1941,floryan1991} for unstable Görtler vortices. These values are however relatively mild compared to cases where transition is dominated by secondary Görtler vortex instability \cite{schrader_brandt_zaki_2011,souza_2017,Xu_Zhang_Wu_2017,li_choudhari_paredes_2022,xu_ricco_duan_2024}.

\section{\label{appE}Comparison with zero-pressure-gradient supersonic boundary layer}

\begin{figure}
    \includegraphics[width=0.8\textwidth]{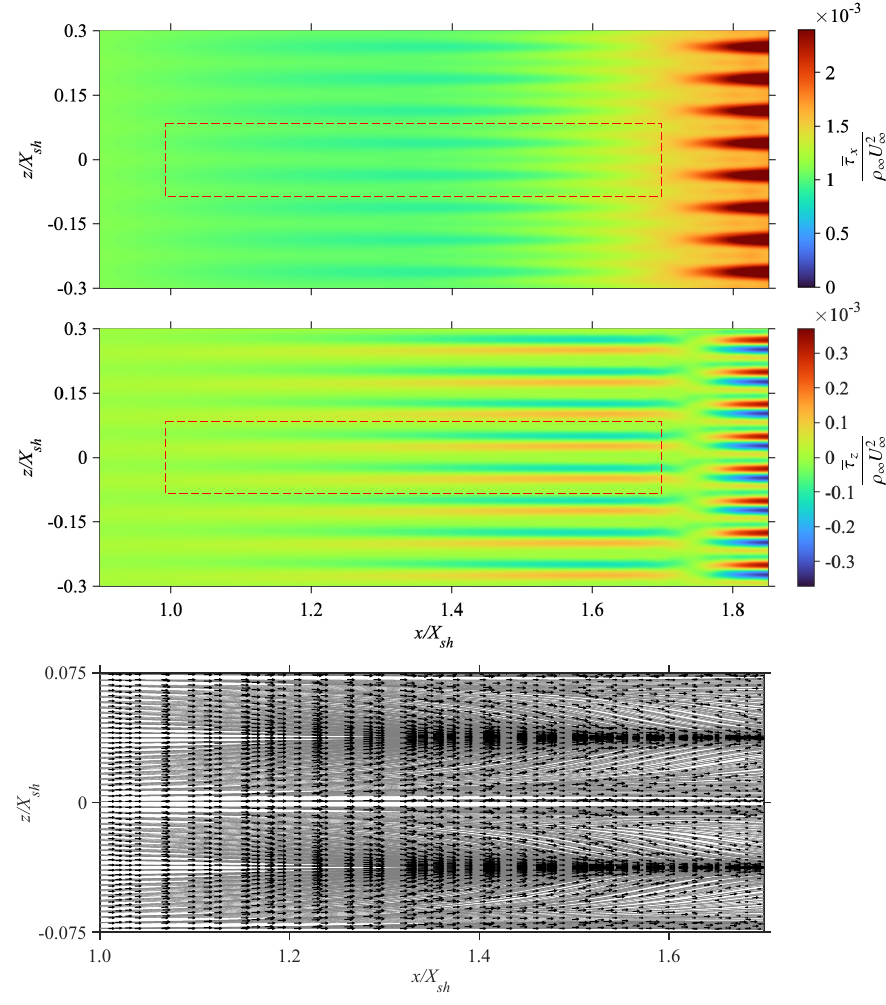}
    \caption{Trace of lift-up mode in the time-averaged flow field of a Mach 2.15 \ac{ZPG} boundary layer at amplitude $A=40.0\times10^{-5}$. Top: time-averaged streamwise wall shear stress. Middle: time-averaged spanwise wall shear stress. Bottom: skin friction lines at the wall and vectors from the wall shear vector field $(\overline{\tau}_x,\overline{\tau}_z)$ plotted within the part of the domain highlighted with the red-dashed box.}
    \label{fig:wall shear ZPG}
\end{figure}

The lift-up-dominated \ac{ZPG} boundary layer time-averaged flow topology is illustrated in figure \ref{fig:wall shear ZPG} and serves as a comparison with the \ac{SWBLI} flow where Görtler-like streamwise vortices are active. The chosen amplitude calculation for the \ac{ZPG} boundary layer yields the same excursion in the values of streamwise and spanwise wall shear stresses to facilitate comparability. Both cases display similar striation patterns in both streamwise and spanwise stresses, but in the \ac{SWBLI} the spanwise component is stronger relative to the streamwise counterpart in the reattachment and post-reattachment zones where the Görtler instability is active. This results in different skin friction line topologies, where lift-up in \ac{ZPG} conditions gives a more gentle bending pattern.

\begin{figure}
    \includegraphics[width=0.40\textwidth]{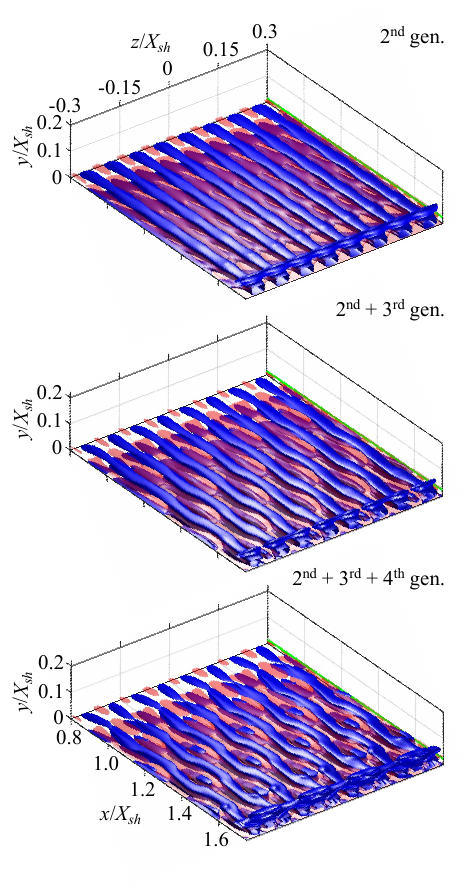}
    \caption{Sub-harmonic sinuous streak instability in supersonic Mach 2.15 \ac{ZPG} boundary layer at high amplitude $A=40.0\times10^{-5}$. The same reconstruction of figure \ref{fig:streak instability} is performed. Mean boundary layer thickness and displacement thickness are shown in green-solid and red-dashed lines, respectively.}
    \label{fig:streak instability ZPG}
\end{figure}

A three-dimensional representation of the secondary streak instability via the sinuous mode under \ac{ZPG} conditions is provided in figure \ref{fig:streak instability ZPG}. In common with figure \ref{fig:streak instability}, the instantaneous $u'$ flow field is reconstructed taking away the 1st generation Mack waves. Streamwise-oriented streaks are reconstructed with the 2nd generation $(2,0)$, $(0,2)$ and $(2,2)$ harmonics (top panel). The flow structures highlight the strong contribution of the streaky $(0,2)$ mode. Spanwise meandering of the low-speed streaks appears when the 3rd generation $(1,3)$ harmonic is superimposed. This mode originates from the non-linear interaction of $(0,2)$ streaks and $(1,1)$ oblique waves, a mechanism thoroughly described in literature \cite{schmid1992new,hanifi1996transient,andersson_brandt_bottaro_henningson_2001,rigas2021HBM,poulain2024}. The addition of 4th generation harmonics emphasizes the sinuous streak pattern.

\bibliography{References_FS2}

\end{document}